\documentclass{aa}  
\usepackage{graphicx}
\usepackage[T1]{fontenc}
\usepackage{txfonts}
\usepackage{placeins}
\usepackage{epstopdf}
\usepackage{newtxtext,newtxmath}
\newcommand{\asec}{$^{\prime\prime}$}

\def\SigmaH2{$\Sigma $(${\rm H_2}$)}
\def\r1415{$^{14}$N/$^{15}$N}

\def\cyclic{{\it c-}C$_3$H$_2$}

\def\H{N$_{2}$H$^{+}$}

\def\15N{$^{15}$NNH$^+$}
\def\N15{N$^{15}$NH$^+$}

\def\HCOp{\mbox{HCO$^+$}}

\def\METH{CH$_3$OH}
\def\FORM{H$_2$CO}

\def\H13CN{\mbox{H$^{13}$CN}}

\def\kms{\mbox{km~s$^{-1}$}}
\def\cmc{cm$^{-3}$}
\def\cmq{cm$^{-2}$}

\def\Tex{\mbox{$T_{\rm ex}$}}

\def\Tk{\mbox{$T_{\rm k}$}}

\def\TMB{\mbox{$T_{\rm MB}$}}

\def\kms{km\,s$^{-1}$}

\begin{document} 

   \title{CHEMical complexity in star-forming regions of the OUTer Galaxy (CHEMOUT). II. Methanol formation at low metallicity}
   
   \author{F. Fontani
          \inst{1,2} 
          \and
          A. Schmiedeke\inst{2}
          \and
          A. S\'anchez-Monge\inst{3}          
          \and
          L. Colzi\inst{4,1}
          \and
          D. Elia\inst{5} 
          \and
          V.M. Rivilla\inst{4,1}
          \and
          M.T. Beltr\'an\inst{1} 
          \and
          L. Bizzocchi\inst{6,2}
          \and
          P. Caselli\inst{2}         
          \and
          L. Magrini\inst{1}
          \and
          D. Romano\inst{7}
          }

   \institute{INAF-Osservatorio Astrofisico di Arcetri, Largo E. Fermi 5, I-50125, Florence, Italy\\
            \email{francesco.fontani@inaf.it}
            \and
            Centre for Astrochemical Studies, Max-Planck-Institute for Extraterrestrial Physics, Giessenbachstrasse 1, 85748 Garching, Germany
            \and
            I. Physikalisches Institut, Universit\"at zu K\"oln, Z\"ulpicher Str. 77, 50937 K\"oln, Germany
            \and
            Centro de Astrobiolog\'ia (CSIC-INTA), Ctra. de Ajalvir Km. 4, Torrej\'on de Ardoz, 28850 Madrid, Spain
            \and
            INAF - IAPS, via Fosso del Cavaliere, 100, I-00133 Roma, Italy
            \and
            Dipartimento di Chimica ''Giacomo Ciamician'', Universit\`a di Bologna, Bologna, Italy
            \and
            INAF, Osservatorio di Astrofisica e Scienza dello Spazio, Via Gobetti 93/3, 40129, Bologna, Italy
             }

 \date{Received ; accepted }

% \abstract{}{}{}{}{} 
% 5 {} token are mandatory
 
 \abstract
   {The outer Galaxy is an environment with metallicity lower than the Solar one and, because of this,
   the formation and survival of molecules in star-forming regions located in the inner and outer Galaxy
   is expected to be different.}
   {To gain understanding on how chemistry changes throughout the Milky Way, it is crucial to observe outer 
   Galaxy star-forming regions to constrain models adapted for lower metallicity environments.}
   {The project "chemical complexity in star-forming regions of the outer Galaxy" (CHEMOUT) aims to
   address this problem observing a sample of 35 high-mass star-forming cores at Galactocentric distances
   up to $\sim 23$~kpc with the Institut de RadioAstronomie Millim\'etrique (IRAM) 30m telescope in various 3mm and 2mm bands.
   In this work we analyse observations of methanol (\METH), one of the simplest complex organic molecules
   crucial for organic chemistry in star-forming regions, and of two chemically related species, HCO and 
   formaldehyde (H$_2$CO), towards 15 out of the 35 targets of the CHEMOUT sample. In fact, only targets 
   previously detected in both HCO and H$_2$CO, which are precursors of \METH, were considered.}
   {We detected \METH\ in all 15 targets. The emission is associated with an extended envelope, as the 
   average angular size is $\sim 47$\asec\ (i.e. $\sim 2.3$~pc at a representative heliocentric distance of 10~kpc). 
   Using a Local Thermodynamic Equilibrium approach, we derive \METH\ excitation temperatures in the range 
   $\sim 7 - 16$ K and line widths $\leq 4$~\kms, consistent with emission from a cold and quiescent envelope. 
   The \METH\ fractional abundances w.r.t. H$_2$ range between 
   $\sim 0.6\times 10^{-9}$ and $\sim 7.4\times 10^{-9}$. These values are comparable to those found in star-forming
   regions in the inner and local Galaxy. \FORM\ and \METH\ show well 
   correlated line velocities, line widths, and fractional abundances w.r.t. H$_2$,
   indicating that their emission is originated from similar gas. These correlations are not seen with HCO,
   suggesting that \METH\ is likely more chemically related to \FORM\ than to HCO. }
   {Our results have important implications in the organic, and possibly pre-biotic, chemistry occurring in
   the outermost star-forming regions of the Galaxy, and can help setting the frontiers of the 
   Galactic Habitable Zone.}


\titlerunning{CHEMOUT II: methanol formation at low metallicity}

\maketitle
%
%-------------------------------------------------------------------

\section{Introduction}
\label{intro}

\begin{table*}
\begin{center}
\label{tab:sources}
\caption{Source list and parameters taken from paper I.}
\begin{tabular}{lcccccccc}
\hline
source &  R.A. & Dec. &  $N_{\rm CO}{\rm (H_2)}$$^{(a)}$ & $N_{\rm Her}{\rm (H_2)}$$^{(b)}$ & $\theta_{\rm c}$$^{(c)}$ & $R_{\rm GC}$$^{(d)}$ & $d$$^{(e)}$ & rms$^{(f)}$  \\
            & (J2000) & (J2000)  &                                            &   &   &                          &    & 3mm/2mm   \\
            &  ${h : m : s}$ & ${\circ : \prime : \prime\prime}$  & $\times 10^{21}$cm$^{-2}$ & $\times 10^{22}$\cmq\ & \asec\ & kpc & kpc & mK  \\
\hline
WB89-379 & 01:06:59.9 & 65:20:51     & 6.5    & 5.76 & 14.6   & 16.4     & 10.2 & 7.5/13.5 \\
WB89-380 & 01:07:50.9 & 65:21:22	   & 11.4 & -- & --       & 16.0  & 9.7 & 7.7/12.1 \\
WB89-391 & 01:19:27.1 & 65:45:44	   & 5.2  & -- & --       & 16.1   & 9.7 & 6.6/11.1  \\
WB89-399 & 01:45:39.4 & 64:16:00     &  6.3   & 4.06 & 34.1   & 16.0    & 9.4 & 6.2/9.7    \\
WB89-437 & 02:43:29.0 & 62:57:08     & 14.2   & -- & --       & 15.7    & 8.6 & 7.7/13.7 \\
WB89-501 & 03:52:27.6 & 57:48:34     & 11.2   & -- & --       & 15.6    & 8.0 & 7.8/14.8 \\
WB89-621 & 05:17:13.4  & 39:22:15     & 13.0  & 14.5 & 11.7   & 18.9   & 10.6 & 9.1/14.8 \\
WB89-789 & 06:17:24.3 & 14:54:37      & 5.8   & 13.7 & 12.4   & 19.1    & 11.0 &  6.8/12.0 \\
19383+2711 & 19:40:22.1 & 27:18:33     & --   & -- & --       & 13.2    &  14.8 & 6.4/9.3  \\
19423+2541 & 19:44:23.2 & 25:48:40     &  --  & 20.1 & 8.3    & 13.5   & 15.3 &  8.1/10.2 \\
WB89-006 & 20:42:58.2 & 47:35:35	  & 6.3   & -- & --       & 14.3    & 12.2 & 4.7/6.7   \\
WB89-035 & 21:05:19.7 & 49:15:59	   & 5.2  & 6.2 & 11.9    & 13.1   & 10.1 &  5.7/11.0 \\
WB89-076 & 21:24:29.0 & 53:45:35	    & 5.0 & 5.6 & 10.5    & 15.1  & 11.8 & 7.8/11.6 \\
WB89-080 & 21:26:29.1 & 53:44:11	    & 8.5 & 2.8 & 18.1    & 12.8  & 8.9 & 7.4/12.6  \\
WB89-283 & 23:32:23.8 & 63:33:18	   & 5.8  & 2.8 & 10.5    & 15.8   & 10.4 & 4.6/6.9  \\
\hline
\end{tabular}
\end{center}
$^{(a)}$ H$_2$ column density derived from CO (1--0) (Blair et al.~\citeyear{blair08}). The values are averaged within the Arizona Radio Observatory (ARO) main beam of 44\asec; \\
$^{(b)}$ H$_2$ column density derived from Herschel measurements (Elia et al.~\citeyear{elia21}). The values are averaged within $\theta_{\rm c}$; \\
$^{(c)}$ continuum angular size estimated from Herschel measurements (Elia et al.~\citeyear{elia21}); \\
$^{(d)}$ Galactocentric distance; \\
$^{(e)}$ heliocentric distance; \\
$^{(f)}$ 1$\sigma$ rms achieved in the IRAM-30m spectra (Appendix-A) around the target \METH\ lines. \\
%, assuming a standard CO-H$_2$
%conversion factor of~1.8$\times 10^{20}$ cm$^{-2}$ (K Km s$^{-1}$)$^{-1}$ (Dame et al.~\citeyear{dame01})
%valid at the Solar Circle.
\end{table*}

The outer Galaxy (OG) is the portion of the Milky Way located beyond the Sun, that is at a distance 
from the Galactic Centre, $R_{\rm GC}$, approximately in between the Solar circle (at $\sim 8.34$~kpc,
Reid et al.~\citeyear{reid14}) and $\sim 28$~kpc (Digel et al.~\citeyear{digel94}). 
Inner and outer Galaxy 
show significantly different chemical properties. First, in the outer Galaxy the overall content of elements 
heavier than helium, defined as metallicity, is lower than the Solar one (e.g.~Wenger et al.~\citeyear{wenger19}). 
The elemental abundances of 
oxygen, carbon, and nitrogen, namely the three most abundant elements in the Universe after hydrogen 
and helium, decrease linearly (in logarithmic scale) as a function of $R_{\rm GC}$ (e.g. Esteban et al.~2017,
Arellano-C\'ordova et al.~\citeyear{arellano20}, M\'endez-Delgado et al.~\citeyear{mendez22}). 
Because of the lower abundances of heavy elements in the outer Galaxy with respect to the Solar ones and those of 
the inner Galaxy, it has been suggested that this zone is not suitable for forming planetary systems 
in which Earth-like planets could be born and might be capable of sustaining life (Prantzos~2008, 
Ram\'irez et al.~2010). The so-called Galactic Habitable Zone (GHZ) in the Milky Way is currently
defined as an annular
region about 2~kpc wide centred at $R_{\rm GC}$~8 kpc, where the metallicity is appropriate to form 
Earth-like planets and where 
the occurrence of disruptive events such as supernovae is limited (Spitoni et al.~2014, 2017). Because 
of this, and also because of the fact that star-forming regions in the outer Galaxy are on average further 
away from us, the study of the formation of stars and planets, as well as the search for the basic bricks 
of life that could have favoured the emergence of life, were focused so far almost exclusively to the inner 
Galaxy, where the metallicity is Solar or super-Solar. 

This scenario has been challenged by recent observational results. First, the occurrence of Earth-like 
planets does not seem to depend on the Galactocentric distance (e.g. Buchhave et al.~\citeyear{buchhave12};
Maliuk \& Budaj~\citeyear{meb20}). This indicates 
that, even at metallicities lower than the Solar one, planets capable to host life can be found,
depending also on the dynamical history of the planet host stars (Dai et al.~\citeyear{dai21}), 
which is likely to be different in the inner and outer Galaxy. 
Second, observations performed with the Atacama Large Millimeter Array (ALMA) towards the Large and Small 
Magellanic Clouds (LMC and SMC, Shimonishi et al.~\citeyear{shimonishi18}, Sewi\l{}o et al.~\citeyear{sewilo18},
Sewi\l{}o et al.~\citeyear{sewilo22}), two external galaxies having metallicity $\sim 3$ and $\sim 5$ times lower
than Solar, respectively, have
detected emission of complex organic molecules (COMs), which are organic species with more than five atoms. 
This finding, and the recent discovery of a hot molecular core rich in COMs at $R_{\rm GC}$ of $\sim 19$~kpc
(Shimonishi et al.~\citeyear{shimonishi21}), have clearly suggested that the astrochemical processes that can 
lead to species of pre-biotic interest can be found also in metal poor environments. In particular, methanol (\METH), 
one of the simplest but crucial COMs, was detected in star-forming regions associated with both the LMC
and SMC, as well as with star-forming regions in the Milky Way up to $R_{\rm GC}\sim 20$kpc (Bernal
et al.~\citeyear{bernal21}). Methanol is thought to be a crucial species for pre-biotic chemistry as it is
considered a possible parent species for larger organic molecules in both gas and ice (e.g.~Charnley et 
al.~\citeyear{charnley92}; \"{O}berg et al.~\citeyear{oberg09}; Chen et al.~\citeyear{chen13}; 
Chuang et al.~\citeyear{chuang16}). Hence, its presence paves the way for the synthesis of more complex 
organic species. 
%, and methyl formate and dimethyl ether was found in hot-cores of the LMC (Sewi\l{}o et al.~\citeyear{sewilo18}). Because these 
%COMs are thought to be precursors of more complex biogenic molecules (see e.g. Caselli \& Ceccarelli~\citeyear{cec12}), 
%these observational findings indicate that the basic bricks of life can be found also in metal poor environments. These
%findings were reinforced by the recent detection of a hot molecular core, rich in COMs, in the extreme outer Galaxy
%(Shimonishi et al.~\citeyear{shimonishi21}).

In this paper we study the emission of methanol, of the formyl radical (HCO) and of formaldehyde 
(H$_2$CO) towards 15 high-mass star-forming regions in the OG with $R_{\rm GC}$ in between
13.1 and 19~kpc. Even though the OG is more extended than 19~kpc (see above),
the $R_{\rm GC}$ range studied in this work is where the metallicity gradients are better constrained
by observations (e.g.~Esteban et al.~\citeyear{esteban17}; Kovtyukh et al.~\citeyear{kovtyukh22}). 
Beyond 20~kpc, the metallicity gradients are much less well constrained (Spina et al.~\citeyear{spina22}).
HCO and \FORM\ are precursors of methanol 
(Watanabe \& Kouchi~\citeyear{wek02}) and of other COMs relevant for pre-biotic chemistry 
such as formamide (e.g.~Fedoseev et al.~\citeyear{fedoseev16}), glycolaldehyde and ethylene glycol 
(Bennett \& Kaiser~\citeyear{bek07}; Woods et al.~\citeyear{woods12},~\citeyear{woods13}; 
Chuang et al.~\citeyear{chuang16}, Rivilla et al.~\citeyear{rivilla17},~\citeyear{rivilla19}). Hence, these 
observations can indicate us for the first time whether the formation pathways of \METH, known to be 
efficient starting from Solar-like metallicities, are efficient also in the lower metallicity environment of the OG.

Our targets are part of CHEMOUT, an observational project described in Fontani
et al.~(\citeyear{fontani22}, hereafter paper I) performed with the Institut de RadioAstronomie
Millim\'etrique (IRAM) 30m telescope\footnote{http://www.iram.es/IRAMES/mainWiki/FrontPage} 
that aims to study the chemical complexity in the low-metallicity environment of the OG. 
Sample and observations are described in Sect.~\ref{sample}. The data analysis is
described in Sect.~\ref{analysis}. The results are presented in Sect.~\ref{res} and discussed in 
Sect.~\ref{discu}.

\section{Sample and observations}
\label{sample}

\begin{table}
\tiny
\label{tab:lines}
\setlength{\tabcolsep}{1.8pt}
\begin{center}
\caption[]{Parameters of the \METH, HCO, and H$_2$CO transitions observed in this work and used in the analysis 
described in Sect.~\ref{analysis}.
}
\begin{tabular}{cccc}
\hline \hline
 $\nu$$^{(1)}$ & Quantum & $E_{\rm u}$$^{(2)}$ & $A_{\rm ul}$$^{(3)}$ \\
 MHz       &   Numbers            &  K  &  $\times 10^{-6}$ s$^{-1}$    \\
\hline
\multicolumn{4}{c}{\METH} \\
\hline
\multicolumn{4}{c}{3~mm} \\
%\hline
96739.358 &  2(1,2)--1(1,1) E$_2$  & 12.5 &  2.5 \\
96741.371 &  2(0,2)--1(0,1) A$^+$   &  7.0  &  3.4 \\
96744.545 &  2(0,2)--1(0,1) E$_1$   & 20.1 &  3.4  \\
96755.501 &  2(1,1)--1(1,1) E$_1$   & 28.0 &  2.6 \\
%\hline
\multicolumn{4}{c}{2~mm} \\
%\hline
%143865.810 & 3(1,3)--2(1,2) A$^+$ & 21.4 & 28.3 & $1.1\times 10^{-5}$ \\
145093.707 & 3(0,3)--2(0,2) E$_1$        & 27.1 & $12$ \\
145097.370 & 3(1,3)--2(1,2) E$_2$        & 19.5 & $11$ \\
145103.152 & 3(0,3)--2(0,2) A$^+$     & 13.9 & $12$ \\
145124.410 & 3(2,2)--2(2,1) A$^-$      & 51.6 &  $6.9$ \\
145126.191 & 3(2,1)--2(2,0) E$_1$    & 36.2 &  $6.8$ \\
145126.386 & 3(2,2)--2(2,1) E$_2$         & 39.8 &  $6.9$ \\
145131.855 & 3(1,2)--2(1,1) E$^1$    & 35.0 & $11$ \\
145133.460 & 3(2,1)--2(2,0) A$^+$  & 51.6 & $6.9$ \\
%146368.328 & 3(1,2)--2(1,1) A         & 21.6 & 28.6 & $1.1\times 10^{-5}$ \\
\hline
\multicolumn{4}{c}{HCO} \\
\hline
%\multicolumn{5}{c}{3~mm} \\
 86670.760          &  $N_{K_a,K_b}=1_{0,1}-0_{0,0}$, $J=3/2-1/2$, $F=2-1$ & 4.2 & $4.7$  \\
 86708.360          &  $N_{K_a,K_b}=1_{0,1}-0_{0,0}$, $J=3/2-1/2$, $F=1-0$  & 4.2 & $4.6$  \\ 
 86777.460          &  $N_{K_a,K_b}=1_{0,1}-0_{0,0}$, $J=1/2-1/2$, $F=1-1$ & 4.2 & $4.6$  \\
 86805.780          &  $N_{K_a,K_b}=1_{0,1}-0_{0,0}$, $J=1/2-1/2$, $F=0-1$ & 4.2 & $4.7$   \\
\hline
\multicolumn{4}{c}{H$_2$CO} \\
\hline
%\multicolumn{5}{c}{2~mm} \\
140839.502        &    $J_{K_a,K_b} = 2_{1,2} - 1_{1,1}$  & 22 & $53$ \\
145602.949         &   $J_{K_a,K_b} = 2_{0,2} -1_{0,1}$   & 10  &  $78$ \\
\hline
\end{tabular}
\end{center}
$^{(1)}$ rest frequencies ($\nu$); $^{(2)}$ energy of the 
upper level ($E_{\rm u}$); $^{(3)}$ Einstein coefficient for spontaneous emission ($A_{\rm ul}$).
%All parameters of 
%\METH\ and H$_2$CO lines are taken from the CDMS (Endres et al.~\citeyear{endres16}), while for the HCO 
%lines they are taken from the Jet Propulsion Laboratory (JPL, Pickett et al.~\citeyear{pickett98}). 
\end{table}
\normalsize

We targeted 15 sources extracted from the sample described in paper I. This sample is made 
of 35 targets selected 
from \citet{blair08}, who searched for formaldehyde emission with the Arizona Radio Telescope (ARO) 
12m telescope in dense molecular cloud cores in the OG. All cores are associated with IRAS
colours typical of star-formation regions.
The targets of this work, listed in Table~\ref{tab:sources}, were selected because they were detected in both 
HCO (paper I) and H$_2$CO (Blair et al.~\citeyear{blair08}), and their $R_{\rm GC}$ cover a wide range 
among the sources detected in HCO (see paper I). In the same Table we give coordinates 
and other source parameters useful for the analysis, such as H$_2$ column densities estimated from 
previous works, and both Galactocentric and heliocentric ($d$) distances estimated in paper I.

The observed lines are listed in Table~\ref{tab:lines}. All parameters of the \METH\ and \FORM\ 
lines are taken from the Cologne Database for Molecular Spectroscopy (CDMS\footnote{https://cdms.astro.uni-koeln.de/classic/}, 
Endres et al.~\citeyear{endres16}), while for the HCO lines they are taken from the Jet Propulsion 
Laboratory(JPL\footnote{https://spec.jpl.nasa.gov/ftp/pub/catalog/catdir.html}, Pickett et al.~\citeyear{pickett98}).

Observations of the \METH\ and H$_2$CO lines listed in Table~\ref{tab:lines} were performed with the IRAM-30m
telescope in 
two observing runs (July and Septembre, 2021; project 042-21). We used the 3 and 2~mm receivers 
simultaneously. We note that 11 of our 15 targets were recently observed in the same \METH\ lines at 
3~mm by \citet{bernal21} with the ARO telescope, and 10 of them were detected. In this work, we 
reobserve these lines with higher sensitivity (rms of $\sim 5-10$~mK versus $\sim 17-200$~mK, 
in main beam temperature units) and higher angular resolution (25\asec\ versus 63\asec). Moreover, we 
add to the analysis the 2~mm lines, which have energies of the upper level up to $\sim 50$~K, and hence 
allow us to constrain the \METH\ parameters, in particular \Tex, more accurately than with the 3~mm 
lines only, for which the range of energy of the upper level is $\sim 7-28$~K (Table~\ref{tab:lines}). 
Sources WB89-006, WB89-789, WB89-035, and WB89-080 have been observed in both bands for the first time. 

The Local Standard of Rest (LSR) velocities used to centre the spectra are listed in Table~1 of paper I. 
The observations were made in wobbler-switching mode with a wobbler throw of 220\asec. 
Pointing was checked (almost) every hour on nearby quasars or bright HII regions. Focus was checked 
on planet Saturn at the start of observations and after sunset. The data were calibrated with the chopper 
wheel technique (see Kutner \& Ulich~\citeyear{keu81}), with a calibration uncertainty of about $10\%$. 
The telescope half power beam width (HPBW) is $\sim 25$\asec\ and $\sim 17$\asec\ in the 3 and 2~mm 
bands, respectively. The spectra were obtained in main beam temperature units 
with the fast Fourier transform spectrometer with channel width 200 kHz (FTS200), providing a 
spectral resolution of~$\sim 0.6$~\kms\ and $\sim 0.4$~\kms\ at 3 and 2~mm, respectively. 
The total bandwidth observed is $90400 - 98180$~MHz and $140720 - 148500$~MHz at 3 and 2~mm,
respectively.
The 1$\sigma$ rms noise level for each spectrum around the \METH\ lines is given in Table~\ref{tab:sources}.

The observations of the HCO lines listed in Table~\ref{tab:lines} were performed 
with the IRAM-30m telescope in the observing runs described in paper I. We
refer to that paper for any observational and technical details related to these data.

\section{Data reduction and analysis}
\label{analysis} 

\begin{table*}
\begin{center}
\label{tab:bestfit-ch3oh}
\caption{\METH\ line parameters.}
\begin{tabular}{lccccccc}
\hline
source & $T^{\rm ARO}_{\rm MB}$$^{(1)}$ & $T^{\rm IRAM}_{\rm MB}$$^{(1)}$ & $\theta_{\rm S}$(\METH)$^{(2)}$  & $V$$^{(3)}$ & FWHM$^{(3)}$ & $N_{\rm tot}$$^{(3)}$ & $T_{\rm ex}$$^{(3)}$ \\
       &      K              &      K             &  arcsec         & \kms\ & \kms\ & $\times 10^{13}$\cmq\ & K \\
 \hline
WB89-379 &     0.04(0.01)   &  0.085  &  43(11)  &  --89.25(0.04) &    1.65(0.10) & 1.4(0.1) & 11.0(1.0) \\
WB89-380 &     0.043(0.007)  &  0.102  &  42(8)  &  --86.48(0.05) &    3.44(0.12) & 3.5(0.2) & 12.0(1.0) \\
WB89-391 &      0.06(0.008)  &  0.133  &  48(9)  &  --85.91(0.03) &    1.48(0.08)  & 1.4(0.1) & 10(1) \\
WB89-399 &     0.023(0.008)  &  0.029  &  110(35)  &  --82.21(0.06) &    1.04(0.10) &  0.25(0.05) & 12(2) \\
WB89-437 &     0.09(0.01)    &  0.240  &  36(6)  &  --71.57(0.03) &    2.98(0.08) & 9.1(0.3) & 14(0.5) \\
WB89-501 &      0.045(0.008) &  0.094  &  51(11)  &  --58.63(0.05) &    1.75(0.1 ) & 1.5(0.1) & 11.0(1.0) \\
WB89-621 &      0.20(0.01)   &  0.473  &   42(5)  &  --25.38(0.03) &    1.74(0.06) & 8.5(0.4) & 12(1) \\
WB89-789 &        --         &  0.081  &      --$^{(4)}$         & 34.06 (0.05)     &    2.10(0.13) & 1.0(0.1)  & 9.7(0.8) \\
19383+2711 &   0.028         &  0.051  &  58(6)  &   --65.25(0.05) &    2.17(0.12) & 1.7(0.1) & 15(1) \\
19423+2541 &   0.049(0.005)  &  0.111  &  45(7)  &   --72.42(0.05) &    3.95(0.12) & 4.1(0.1) & 12.2(0.5) \\
WB89-006 &        --         &  0.108  &      --$^{(4)}$         &  --91.37(0.05) &    2.92(0.12)  & 1.9(0.1)  & 9.6(0.5) \\
WB89-035 &        --         &  0.045  &      --$^{(4)}$         &  --77.70(0.07) &    2.28(0.16) & 1.1(0.8) & 10.9(0.6) \\
WB89-076 &     0.087(0.008)  &  0.170  &  54(8)  &  --97.23(0.04) &    1.56(0.10) & 1.5(0.2)  & 7.0(0.5) \\
WB89-080 &        --         &  0.081  &      --$^{(4)}$         &  --74.41(0.06) &    1.89(0.13) & 1.0(0.1)  & 9.8(0.9) \\
WB89-283 &        --         &  0.058  &      --$^{(4)}$         &  --94.45(0.03) &    1.31(0.07) & 0.83(0.05) & 16.4(1.4) \\
\hline
\end{tabular}
\end{center}
$^{(1)}$ peak intensity of the 2(0,2)--1(0,1) A$^+$ transition
observed with the ARO-12m ($T^{\rm ARO}_{\rm MB}$) and IRAM-30m ($T^{\rm IRAM}_{\rm MB}$) telescope, respectively.
$T^{\rm ARO}_{\rm MB}$ was derived from the $T^*_{\rm a}$ units given in \citet{bernal21} as described in Sect.~\ref{analysis};
on $T^{\rm IRAM}_{\rm MB}$, a calibration uncertainty of $10\%$ is assumed; \\
$^{(2)}$ angular size of the emission derived as explained in Sect.~\ref{analysis}; \\
$^{(3)}$ best fit parameters of the \METH\ lines obtained with {\sc madcuba} fixing the source size
to the value in Col.~4: centroid velocity ($V$), 
full width at half maximum (FWHM), total column density ($N_{\rm tot}$), and excitation temperature ($T_{\rm ex}$); \\
$^{(4)}$ an average size of 47\asec\ has been assumed to fit the data. \\
%, assuming a standard CO-H$_2$
%conversion factor of~1.8$\times 10^{20}$ cm$^{-2}$ (K Km s$^{-1}$)$^{-1}$ (Dame et al.~\citeyear{dame01})
%valid at the Solar Circle.
\end{table*}
\normalsize

The first steps of the data reduction (e.g. average of the scans, baseline removal, flag of 
bad scans and channels) were made with the {\sc class} package of the 
{\sc gildas}\footnote{https://www.iram.fr/IRAMFR/GILDAS/} software using standard procedures.
Then, the baseline-subtracted spectra in main beam temperature ($T_{\rm MB}$) units were 
fitted with the MAdrid Data CUBe Analysis ({\sc madcuba}\footnote{{\sc madcuba} is a software developed in the Madrid Center of Astro-biology (INTA-CSIC) which enables to visualise and analyse single spectra and data cubes: https://cab.inta-csic.es/madcuba/.}, Mart\'in et al.~\citeyear{martin19}) software.

The transitions of HCO, H$_2$CO, and \METH\ in the bands described in Sect.~\ref{sample}
were identified via the SLIM (Spectral Line Identification and 
LTE Modelling) tool of {\sc madcuba}. The lines were fitted with the AUTOFIT function of SLIM. This function 
produces the synthetic spectrum that best matches the data assuming a constant excitation temperature
(\Tex) for all transitions. The other input parameters are: total molecular column density 
($N_{\rm tot}$), radial systemic velocity of the source ($V$), line full-width at half-maximum (FWHM), 
and angular size of the emission ($\theta_{\rm S}$). AUTOFIT assumes that $V$, FWHM and $\theta_{\rm S}$
are the same for all transitions. These input parameters have all been left free except $\theta_{\rm S}$. In
fact, $\theta_{\rm S}$ can be computed for the \METH\ and \FORM\ lines observed also with the
ARO-12m telescope by comparing the line intensities obtained with the two telescopes. 

As said in Sect.~\ref{sample}, \citet{bernal21} detected the same 3~mm \METH\ transitions in 10 of our 15 targets. 
Assuming that the brightness temperature distribution is Gaussian, as well as the beam of the two telescopes, and
that there is no contamination from other sources when moving from the smaller to the larger beam,
one finds that the angular size of the emission is given by (see also Eqs. (2) and (3) in Fontani et al.~\citeyear{fontani02}):

\begin{equation}
\label{eq:size}
\theta_{\rm S} = \frac{\Theta^2_{\rm ARO} - \Theta^2_{\rm IRAM}(T^{\rm IRAM}_{\rm MB}/T^{\rm ARO}_{\rm MB})}{(T^{\rm IRAM}_{\rm MB}/T^{\rm ARO}_{\rm MB})-1}\;,
\end{equation}
where $\Theta_{\rm ARO}$ and $\Theta_{\rm IRAM}$ are the half power beam widths of the two
telescopes at the frequency of the observed lines, and $T^{\rm ARO}_{\rm MB}$ and $T^{\rm IRAM}_{\rm MB}$
are the main beam brightness temperature at line peak obtained with the ARO-12m and IRAM-30m telescope,
respectively. We compare the intensity of the strongest transition, namely the
2(0,2)--1(0,1) A$^+$ one at $\sim 96741$~MHz (Table~\ref{tab:lines}). 
 $\Theta_{\rm ARO}$ and $\Theta_{\rm IRAM}$ are 63\asec\ and 25\asec, respectively, at $\sim 97$~GHz. For each source,
$T^{\rm IRAM}_{\rm MB}$ was obtained by fitting the line with a single Gaussian in {\sc class}, and $T^{\rm ARO}_{\rm MB}$
was obtained by converting the line intensity given in Table~3 of \citet{bernal21} to \TMB\ units by 
dividing it by a factor 0.61. This conversion factor was taken from Appendix C.3 of the ARO-12m user 
manual\footnote{https://aro.as.arizona.edu/$\sim$aro/12m\_docs/12m\_userman.pdf}.
$T^{\rm ARO}_{\rm MB}$, $T^{\rm IRAM}_{\rm MB}$, and the associated $\theta_{\rm S}$ are reported in Table~\ref{tab:bestfit-ch3oh}.
For the sources for which the second strongest \METH\ line is detected in both works, that is the 2(1,2)--1(1,1) E$_2$ 
transition at $\sim 96739$~MHz, we find angular sizes consistent within the uncertainties.

We performed the same analysis for the $J_{K_a,K_b} = 2_{1,2} - 1_{1,1}$ transition of \FORM\ at $\sim 140840$~MHz, 
observed both in this work and in \citet{blair08}. To convert the line intensities given in Table~1 of \citet{blair08} to 
\TMB\ units we divided their values by 0.8\footnote{https://aro.as.arizona.edu/$\sim$aro/12m\_docs/12m\_userman.pdf}.
$\Theta_{\rm ARO}$ and $\Theta_{\rm IRAM}$ are 44\asec\ and 17\asec, respectively, at $\sim 140$~GHz.
The obtained $\theta_{\rm S}$, and the intensities of the lines used to derive them, are listed in Table~\ref{tab:bestfit-h2co}.

\begin{figure}
{\includegraphics[width=8cm]{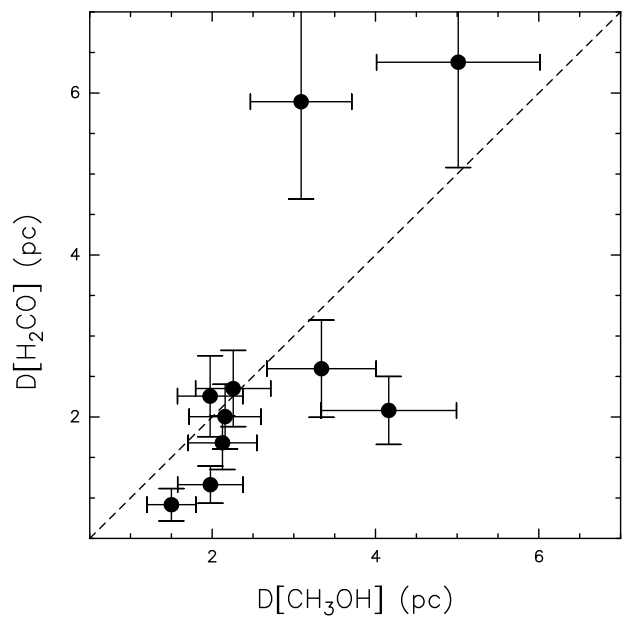}}
      \caption{Comparison between the linear diameters obtained for \METH\ (x-axis) and \FORM\ (y-axis)
      calculated from the angular diameters given in Tables~\ref{tab:bestfit-ch3oh} and ~\ref{tab:bestfit-h2co},
      respectively, and the source heliocentric distances in Table~\ref{tab:sources}. The dashed line is the locus
      where $D$[\METH] = $D$[\FORM].
      }
\label{fig:diameters}
\end{figure}

For both molecules, in all sources $\theta_{\rm S}$ is larger (in some cases much larger) than the IRAM-30m beam. 
This indicates that the observed transitions trace an extended envelope of the cores. In fact, from the heliocentric
distances given in Table~\ref{tab:sources}, we derive that the linear size of the emitting region is $\sim 1.5 - 5$~pc 
for \METH\ and $\sim 0.9 - 6$~pc for \FORM. 
In Fig.~\ref{fig:diameters}, we plot the linear diameter, $D$, of \FORM\ against that of \METH, from which we
see that $D$[\METH] and $D$[\FORM] are positively correlated. There are no systematic similarities or
differences because $D$[\METH] is equal to $D$[\FORM] within the errors in three sources, larger in five, and smaller
in two. Hence, this comparison indicates that the two tracers are associated with extended envelopes of comparable 
dimensions within a factor 2.

In the analysis performed with {\sc madcuba}, for the five sources for which $\theta_{\rm S}$ of \METH\ could not be 
derived from the data, we have assumed the average angular size of the other sources (i.e. 47\asec). 
We have adopted this simplified approach because the five targets have similar heliocentric distances 
(in the range 8.9 -- 12.2~kpc). However, we have checked how the results would change adopting for each source
the angular size obtained from the average linear diameter computed from the sources with measured angular sizes, 
that is $\sim 2.8$~pc. Repeating the fits fixing 
$\theta_{\rm S}$ to these angular sizes, we obtain best fit results consistent within the uncertainties with the results 
obtained from the average angular size for both \Tex\ and $N_{\rm tot}$.
For HCO, because the size of the emission is unknown in the studied lines, we have assumed 
that the emission fills the telescope beam. The assumption is justified by the fact that the energies of the
upper level of the observed transitions are very low ($\sim 4$~K, Table~\ref{tab:lines}), and that HCO is found 
to trace the extended envelope of star-forming cores (e.g.~Rivilla et al.~\citeyear{rivilla19}).

The best fit parameters obtained for \METH\ are listed in Table~\ref{tab:bestfit-ch3oh}. Those obtained for 
H$_2$CO and HCO are given in Table~\ref{tab:bestfit-h2co} and ~\ref{tab:bestfit-hco}, respectively.
For HCO, because the 4 observed lines have the same energies of the upper level, \Tex\ could not be derived
from the data. Hence, we had to fix \Tex. We will discuss the assumed \Tex\ for HCO in Sect.~\ref{hco}.
%We also tried to fix \Tex\ to the values obtained from \FORM\ but the fit either 
%did not converge or gave higher residuals.

\section{Results}
\label{res}

\begin{table*}
\label{tab:bestfit-h2co}
\setlength{\tabcolsep}{2pt}
\begin{center}
\caption[]{\FORM\ line parameters.}
\begin{tabular}{cccccccccc}
\hline \hline
source    & $T^{\rm ARO}_{\rm MB}$$^{(1)}$ & $T^{\rm IRAM}_{\rm MB}$$^{(1)}$ & $\theta_{\rm S}$(\FORM)$^{(2)}$ & $V$$^{(3)}$ & FWHM$^{(3)}$ & $N_{\rm tot}$$^{(3)}$ & $T_{\rm ex}$$^{(3)}$ & $X_{\rm CO}$[\FORM]$^{(4)}$   & $X_{\rm Her}$[\FORM]$^{(5)}$  \\
          &      K             &      K             &  arcsec         & \kms\ & \kms      & $\times 10^{13}$ \cmq\  & K & $\times 10^{-9}$ \cmq\ & $\times 10^{-9}$ \cmq\  \\
\hline
    WB89-379 &   0.255   & 0.54 & 34(3)    & --89.39(0.01) & 1.75(0.04) &    1.20(0.07) & 31(2) & 1.1(0.3) & 1.2(0.3) \\           
    WB89-380 &  0.5713   & 0.93 & 48(5)    & --86.65(0.04) & 3.72(0.08) &     3.8(0.3) & 28(3) & 4(1) & -- \\
    WB89-391 &  0.3463   & 0.55 &  50(5)    & --86.03(0.02) & 1.55(0.04) &    0.85(0.06) & 25(2) & 2.1(0.6) & -- \\
    WB89-399 &  0.4425   & 0.48 & 140(14)    & --82.16(0.02) & 1.51(0.04) &     0.6(0.2) & 26(2) & 10(2) & 2.6(0.6) \\
    WB89-437 &    0.57   & 1.80 & 22(2)    & --71.42(0.02) & 2.72(0.04) &     9.3(0.5) & 33(2) & 1.6(0.4) & -- \\
    WB89-501 &  0.3075   & 0.73 & 30(3)    & --58.46(0.02) & 1.95(0.05) &     2.1(0.2) & 33(4) & 0.9(0.3) & -- \\
    WB89-621 &  0.7863   & 1.50 & 39(4)    & --25.33(0.03) & 2.35(0.07) &     4.0(0.1) & 26(2) & 2.4(0.5) & 3.1(0.7) \\
    WB89-789 &  0.2175   & 0.66 & 23(2)    & 34.20 (0.03) & 3.16(0.07) &     3.2(0.8) & 45(9) & 1.5(0.7) & 0.8(0.4)  \\
  19383+2711 &  0.2687   & 0.67 & 29(3)    & --65.79(0.02) & 2.60(0.06) &     2.9(0.2) & 36(5) & -- & -- \\
  19423+2541 &  0.4813   & 1.00 &  35(4)    & --72.59(0.02) & 3.92(0.06) &     6.5(0.6) & 40(3) & -- & 6(2) \\
    WB89-006 &  0.2037   & 0.33 & 49(5)    & --91.29(0.07) & 3.2 (0.2) &    1.1(0.1) & 26(4) & 2.2(0.7) & -- \\
    WB89-035 &  0.2188   & 0.45 & 36(4)    & --77.58(0.02) & 2.35(0.06) &     1.4(0.1)& 32(3) & 1.8(0.5) & 2.1(0.6) \\
    WB89-076 &  0.3038   & 0.35 & 103(10)    & --97.27(0.03) & 1.84(0.06) &    0.65(0.06) & 28(3) & 7(2) & 11(3) \\
    WB89-080 &  0.2387   & 0.53 & 33(3)    & --74.47(0.03) & 1.65(0.07) &    1.1(0.2) & 30(5) & 0.7(0.3) & 1.3(0.5) \\
    WB89-283 &  0.2625   & 0.53 & 36(4)    & --94.46(0.02) & 1.56(0.04) &    1.2(0.1) & 35(5) & 1.4(0.4) & 5(2) \\              
\hline
\end{tabular}
\end{center}
$^{(1)}$ peak intensity of the $J_{K_a,K_b} = 2_{1,2} - 1_{1,1}$ transition at $\sim 140840$~MHz
observed with the ARO-12m ($T^{\rm ARO}_{\rm MB}$) and IRAM-30m ($T^{\rm IRAM}_{\rm MB}$) telescope, respectively.
$T^{\rm ARO}_{\rm MB}$ was derived from the $T^*_{\rm r}$ units given in \citet{blair08} as described in Sect.~\ref{analysis}.
$T^*_{\rm r}$ is given without uncertainty in \citet{blair08}.
On $T^{\rm IRAM}_{\rm MB}$, a calibration uncertainty of $10\%$ is assumed; \\
$^{(2)}$ angular size of the emission derived as explained in Sect.~\ref{analysis}; \\
$^{(3)}$ best fit parameters of the \METH\ lines obtained with {\sc madcuba} fixing the source size
to the value in Col.~4: centroid velocity ($V$), 
full width at half maximum (FWHM), total column density ($N_{\rm tot}$), and excitation temperature ($T_{\rm ex}$); \\
$^{(4)}$ fractional abundances w.r.t. H$_2$ computed from $N_{\rm CO}$(H$_2$), given in Table~\ref{tab:sources}; \\
$^{(5)}$ fractional abundances w.r.t. H$_2$ computed from $N_{\rm Her}$(H$_2$), given in Table~\ref{tab:sources}. \\
\end{table*}
\normalsize

\subsection{Methanol}
\label{methanol}

We have detected \METH\ emission in both observing bands towards all targets. The observed spectra and 
their best fits (superimposed on them) are shown in Figs.~\ref{fig:spectra-Fig1} and \ref{fig:spectra-Fig2}.
The residuals, shown in Fig.~\ref{fig:residuals}, are generally lower than, or comparable to, the 3$\sigma$ 
rms noise (Table~\ref{tab:sources}) towards all the methanol lines except for five targets that are WB89-391, 
WB89-437, WB89-621, WB89-006, and WB89-076, in which at least one \METH\ line is significantly 
underestimated by the best fit.
This cannot be due to (large) optical depth effects because the best fit opacities provided by {\sc madcuba} 
are always below 0.1. 
However, considering that we assume a constant \Tex\ for all transitions, this can be due to marginal deviations 
from this simplified approach. As stated in Sect.~\ref{sample}, 11 out of our 15 targets 
were already observed in \METH\ at 3~mm by \citet{bernal21} with the ARO telescope, and 10 of them
were detected at 3~mm. In this work, we confirm all their detections and, thanks to our higher sensitivity, 
we confirm their tentative detection of \METH\ claimed towards WB89-283.  Moreover,
we detect also the fainter 3~mm transitions at rest frequencies 96744.545 and 96755.501~MHz (Table~\ref{tab:lines}), 
observed but undetected towards some targets in \citet{bernal21}.

We find \Tex\ in the range $7-16.4$~K, and FWHM of the lines in the range $\sim 1-4$~\kms\
(Table~\ref{tab:sources}). These values confirm what already suggested by \citet{bernal21}, and by 
the estimated angular sizes,
namely that the emission is dominated by the cold and (relatively) quiescent gaseous envelope of the 
cores. This result is also in agreement with the low excitation energy of the 
transition from which $\theta_{\rm S}$ is estimated ($E_{\rm u}= 7$~K, see Table~\ref{tab:lines}).
%Comparable low FWHM, consistent 
%with our results within the uncertainties, were obtained by \citet{bernal21} 
%using only the 3~mm \METH\ transitions listed in Table~\ref{tab:lines}. 
\citet{bernal21} also measured
the kinetic temperatures from a non-LTE analysis, and derived values in the range 10--25~K. This
would indicate that the methanol emission could be sub-thermally excited in several sources. It is
not trivial to deduce if, and by how much, the methanol lines are sub-thermally excited, because
the calculation of the critical densities of the \METH\ lines is not straightforward. In fact, as discussed 
by \citet{shirley15}, the "typical" expression that includes only the Einstein coefficient 
for spontaneous emission and the collisional coefficient is valid only in a two-level approximation, 
not appropriate for molecules with spectrum as complex as that of methanol.

We have used RADEX on-line\footnote{http://var.sron.nl/radex/radex.php} to test possible differences 
with respect to a non-LTE approach, especially for the sources for which our fitting procedure gives the 
highest residuals. For example, the intensity of the 3~mm lines in WB89-621 belonging to the $-A$ and
$-E$ species can be reproduced assuming a kinetic temperature, \Tk, of 15~K, line width equal to the 
observed one (1.74~\kms, Table~\ref{tab:bestfit-ch3oh}), and a H$_2$ volume density, $n$(H$_2$), 
of $3 \times 10^{6}$~\cmc. \Tk\ and $n$(H$_2$) are similar to the values obtained
by \citet{bernal21} with their non-LTE approach. The resulting \Tex\ and total column density, $N_{\rm tot}$,
(derived as the sum of the column density of the $-A$ and $-E$ species) are 14~K and $6\times 10^{13}$~\cmq,
respectively, consistent with the estimates obtained with our approach 
(\Tex$\sim 12$~K and $N_{\rm tot}\sim 8.5\times 10^{13}$~\cmq).
The excitation temperatures and column densities are both comparable for the $-A$ and $-E$ species.

We note that a $n$(H$_2$) of $\sim 10^{6}$~\cmc\ could not be appropriate for the angular scales
we are probing. In fact, from the H$_2$ column densities and diameters provided in Table~\ref{tab:sources}
we derive for WB89-621 a $n$(H$_2$) of $\sim 3\times 10^3$\cmc\ when smoothed on the methanol 
source size of 42\asec. We have then
fixed in RADEX $n$(H$_2$) to $\sim 3\times 10^3$\cmc, and again the FWHM to that observed
and \Tk $\sim 15$~K. The line intensities can be reproduced with $N_{\rm tot}\sim 1.3\times 10^{14}$\cmq\
($7.2\times 10^{13}$\cmq\ for the $-A$ species, $6\times 10^{13}$\cmq\ for the $-E$ species), higher but still roughly
consistent (within a factor 1.5) with the estimate found in the LTE approach also in this case. 
The excitation temperature is now $\sim 4$~K, indicating sub-thermal excitation. However, even with different 
combinations of \Tk, $n({\rm H_2})$, and
\Tex, $N_{\rm tot}$ must always be close to the value obtained in the LTE approach to reproduce the
observed line intensities. 
Similar results can be obtained towards WB89-391, WB89-437, WB89-006, and WB89-076, 
for which a non-LTE approach considering $n({\rm H_2})$ in the range $10^3 - 10^4$\cmc\ always 
needs values of $N_{\rm tot}$ consistent with the results obtained in LTE to reproduce the line intensities. 
In general, the non-LTE approach does not improve significantly the intensity ratios between $-A$ and $-E$ 
lines with respect to the LTE approach, except when using the unrealistic option of a different H$_2$ 
volume density for $-A$ and $-E$ species. Furthermore, the excitation temperatures for $-A$ and $-E$ 
species look consistent each other, and even assuming $-A$ and $-E$ ratios smaller than one does not 
improve the fit results significantly either.

%first, the actual average volume density of the emitting gas is likely closer to the critical density of
%the \METH\ lines, because several targets observed with Herschel show
%angular sizes as small as $\sim 10$\asec\ (Table~\ref{tab:abundances}). In such
%reduced angular sizes, the average H$_2$ volume density would be higher and hence
%closer to the critical density of the analysed lines. 
This, and the general good agreement between the fits and the observed spectra (Figs.~\ref{fig:spectra-Fig1}, 
\ref{fig:spectra-Fig2}, \ref{fig:residuals}), shows that our LTE analysis provides
accurate total column densities, even if some transitions are likely sub-thermally excited.

Shimonishi et al.~(\citeyear{shimonishi21}) found that WB89-789, one
of our targets, harbours a hot core in which the kinetic temperature measured with ALMA 
is higher than 100~K at scales $\leq 0.03$~pc, obtained from \METH\ lines. However, Shimonishi 
et al.~(\citeyear{shimonishi21}) estimated \Tex\ mostly from transitions with $E_{\rm u}\geq 100$~K,
i.e. certainly arising from the inner hot core. 
We have produced with {\sc madcuba} a synthetic spectrum fixing \Tex\ to 100~K, $\theta_{\rm S}$ 
to 0.6\asec\ (corresponding to 0.03~pc at the source heliocentric distance of 11~kpc), and the column
density to the value found by \citet{shimonishi21}, i.e.~$\sim 1.9\times 10^{16}$\cmq. As expected, the 
resulting spectrum is within the noise level of our data, and hence does not give a significant/detectable 
contribution to the emission observed with the IRAM-30m telescope. This confirms that our observations
trace only the external envelope of the hot core, and that in this and potentially also in the other targets
higher angular resolution observations are needed to study the warmer more compact gaseous components.
%Similarly, Sewi\l{}o et al.~(\citeyear{sewilo18}) derived
%methanol temperatures of $\sim 130$~K in hot cores of the LMC again from
%\METH\ lines with excitation energies typically much larger than 100~K.

The \METH\ total column densities obtained towards our targets are in the range $\sim 0.25\times 10^{13}$ 
and $\sim 8.5\times 10^{13}$ cm$^{-2}$ (Table~\ref{tab:bestfit-ch3oh}). These values are consistent within 
a factor $\sim 2$ with those estimated by \citet{bernal21} in the common targets, once a beam dilution factor 
to our values to match their beam of 63\asec\ is applied.  
 
\subsection{Formaldehyde}
\label{formaldehyde} 

\begin{figure}
{\includegraphics[width=11cm]{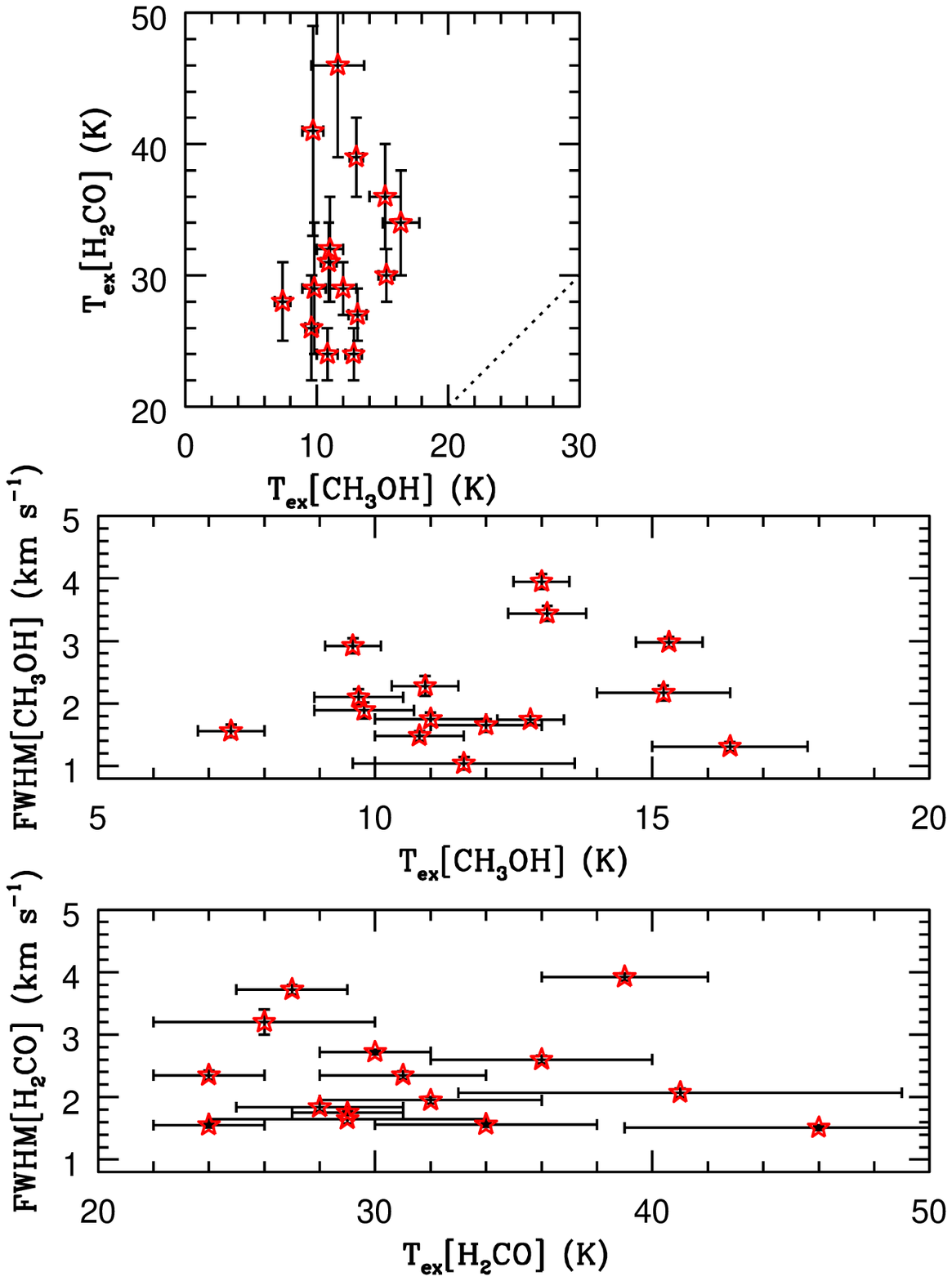}}
      \caption{{\it Top panel}: comparison between the best fit \Tex\ estimated from \FORM\ and \METH.The dotted
      line indicates \Tex[\FORM] = \Tex [\METH].
      {\it Middle and bottom panels:} \Tex\ against FWHM measured from \FORM\ (middle) and from \METH\ (bottom).
      The \METH\ and \FORM\ parameters are given in Tables~\ref{tab:bestfit-ch3oh} and~\ref{tab:bestfit-h2co}, respectively.
      }
\label{fig:temperatures}
\end{figure}

The two formaldehyde lines listed in Table~\ref{tab:lines}, fitted and analysed as described in Sect.~\ref{analysis},
are all clearly detected, as seen in the spectra of Figs.~\ref{fig:spectra-H2CO-Fig1} and \ref{fig:spectra-H2CO-Fig2}.
The line at $\sim 140.8$~GHz was already detected in all sources by \citet{blair08} with an angular resolution
of 44\asec. We confirm all their detections and add in the analysis the line at $\sim 145.6$~GHz.
Some targets present hints of non-Gaussian high-velocity wings in the spectra of both transitions: 
WB89-437, WB89-621, 19423+2541, and WB89-080.
All these sources were known to have high-velocity wings in the \HCOp\ $J=1-0$ line (paper I). A second 
velocity feature towards 19383+2711, already found in \HCOp, is clearly detected also in H$_2$CO.
Two targets, WB89-380 and WB89-006, show hints of two peaks in the $\sim 140.8$ GHz line. In WB89-380,
such profile is not seen in the other line at $\sim 145.6$~GHz, suggesting that it could be due to self-absorption. 
Instead, a similar profile is seen towards the other transition in WB89-006, suggesting that in this case a second 
velocity feature could also be responsible for the tentative second velocity peak. In all cases, however, the
residuals of the Gaussian fits are low.

The best fit parameters are given in Table~\ref{tab:bestfit-h2co}. 
The total column densities are in the range $0.6-9.3\times 10^{13}$~\cmq\ similar to the \METH\ ones. 
The excitation temperatures are 
in the range 25 -- 45~K, i.e. significantly larger than those measured in \METH\ (Table~\ref{tab:bestfit-ch3oh}). 
We will discuss this result more accurately in Sect.~\ref{temp}.

\subsection{Formyl radical}
\label{hco}

The four HCO transitions listed in Table~\ref{tab:lines} were fitted and analysed as described in Sect.~\ref{analysis}.
The observed spectra and the best fit superimposed on them are shown in Figs.~\ref{fig:spectra-HCO-Fig1} and
\ref{fig:spectra-HCO-Fig2}. In most sources, three out of the four transitions are clearly detected at a 
significance level of $\sim 3\sigma$~rms or higher, while that at 86.80578~GHz, having the lowest
line strength, is clearly detected only towards WB89-380 and WB89-391.

The best LTE fit parameters are given in Table~\ref{tab:bestfit-hco}. As said in Sect.~\ref{analysis}, \Tex\
could not be derived from the observations because the energies of the transitions were too close.
Thus, as first order approach, we fixed \Tex\ to the values obtained from \METH.
We obtained FWHM in between 1.15 and 6.5~\kms, although the latter value, obtained towards 19383+2711,
is very likely affected by the presence of a second unresolved velocity feature (clearly detected in \HCOp\ and 
\cyclic, see paper I). Excluding this case, 
the measured FWHM are in between 1.15 and 3.3~\kms, similar to those measured in \METH. 
The HCO total column densities are in the range $0.6 - 9.3\times 10^{12}$~\cmq, i.e. an 
order of magnitude lower than those of \METH\ and \FORM.

An alternative choice for \Tex\ would be that computed from \FORM, which is likely closer to LTE conditions, as
demonstrated by the good agreement between the LTE fits and the spectra (see Figs.~\ref{fig:spectra-H2CO-Fig1} 
and \ref{fig:spectra-H2CO-Fig2}).
However, as we will discuss in Sect.~\ref{temp}, very likely HCO is associated with an envelope more extended
and less turbulent than that traced by both \FORM\ and \METH, based on the significantly narrower line widths at 
half maximum.
The excitation temperature of \FORM\ thus likely represents the gas kinetic temperature of a region 
more turbulent (and likely more compact) than that traced by HCO. Therefore, the smaller \Tex\ derived 
from the (likely) sub-thermally excited \METH\ lines could be closer to the real excitation conditions of HCO. In summary, 
it is not obvious to decide "a priori" which of the two \Tex\ is most appropriate, and thus we decided to fit the HCO lines
also fixing \Tex\ to that of \FORM. The results are given in Table~\ref{tab:bestfit-hco-h2co} of Sect.~\ref{appb}, and 
the alternative fits are shown in Figs.~\ref{fig:spectra-HCO-Fig1-h2co} and \ref{fig:spectra-HCO-Fig2-h2co}. 

With respect to the results performed using the \Tex\ of \METH\ (Table~\ref{tab:bestfit-hco}), line widths at half 
maximum and peak velocities are almost identical, while $N_{\rm tot}$ are systematically higher by a moderate 
factor ($\leq 3$) for all sources except for WB89-789 and 19423+2541, for which they are 
higher by a factor $\sim 5.6$ and $\sim 3.8$, respectively. Because the alternative fits do not seem to reproduce
the observed spectra better than the first order ones (compare Figs.~\ref{fig:spectra-HCO-Fig1} -- \ref{fig:spectra-HCO-Fig2} 
with Figs.~\ref{fig:spectra-HCO-Fig1-h2co} 
-- \ref{fig:spectra-HCO-Fig2-h2co}), it is not obvious to decide which approach is more accurate. In the following 
sections, we will consider the results derived from \Tex\ of \METH, which we believe to better
represent an envelope more extended than that traced by \FORM, bearing in mind that they 
have to be considered as lower limits if the actual \Tex\ of HCO are higher.

\section{Discussion}
\label{discu}

\begin{table*}
\label{tab:bestfit-hco}
\begin{center}
\caption[]{HCO line parameters. }
\begin{tabular}{ccccccc}
\hline \hline
     source        & $V$$^{(1)}$   & FWHM$^{(1)}$      & $N_{\rm tot}$$^{(1)}$       & $T_{\rm ex}$$^{(2)}$ & $X_{\rm CO}$[HCO]$^{(3)}$        & $X_{\rm Her}$[HCO]$^{(4)}$  \\
              & \kms\ & \kms      & $\times 10^{12}$ \cmq\  & K                    & $\times 10^{-10}$ & $\times 10^{-10}$ \\
\hline
    WB89-379 & --89.17(0.09) & 2.6(0.2)  &   2.4(0.2)  & 11.0 & 3.7(0.6)  & 3.8(0.6) \\           
    WB89-380 & --86.47(0.06) & 3.3(0.1) &   6.9(0.3)  & 12.0   & 6.1(0.8) & --       \\
    WB89-391 & --85.94(0.04) & 1.7(0.1)  &   3.0(0.1)  & 10.0  & 5.8(0.8) & --       \\
    WB89-399 & --81.79(0.09) & 2.0(0.2)  &  4.0(0.4)  & 12.0  & 6(1)  & 1.6(0.3) \\
    WB89-437 & --71.8(0.2)    & 2.8(0.4)  &   2.2(0.3)  & 14.0 & 1.6(0.4) & --       \\
    WB89-501 & --58.32(0.07) & 2.0(0.2)  &   3.1(0.2)  & 11.0  & 2.8(0.5) & --       \\
    WB89-621 & --25.5(0.1)   & 2.0(0.2)  &   2.0(0.2)  & 12.0  & 1.5(0.3) & 1.9(0.4) \\
    WB89-789 & 34.21(0.08)  & 2.1(0.2)  &   3.2(0.3)  & 9.7    & 5(1) & 2.9(0.5) \\
  19383+2711 & --68.6(0.2)   & 6.5(0.3)  &   9.3(0.4)  & 15.2  & --       & --       \\
  19423+2541 & --72.53(0.09) & 3.3(0.2)  &  5.0(0.3)  & 12.2   & --       & 7(1) \\
    WB89-006 & --90.5(0.1)    & 1.4(0.3)  &   0.9(0.2)  & 9.6  & 1.4(0.4) & --       \\
    WB89-035 & --77.61(0.07) & 1.15(0.16) &   1.3(0.2)  & 10.9 & 2.4(0.5) & 2.8(0.6) \\
    WB89-076 & --97.1(0.1)  & 1.4(0.2)  &   0.6(0.1)  & 7.0  & 1.3(0.3)   & 2.0(0.5) \\
    WB89-080 & --74.0(0.1)  & 1.4(0.3)  &   1.4(0.3)  & 9.8  & 1.7(0.5)   & 3.0(0.8) \\
    WB89-283 & --94.4(0.14) & 2.5(0.3)  &  2.5(0.3)  & 16.4  & 4(1)   & 16(4) \\         
\hline
\end{tabular}
\end{center}
$^{(1)}$ Best fit parameters obtained with {\sc madcuba}. We assumed that the emission fills the telescope beam; \\
$^{(2)}$ fixed to the value obtained from \METH\ (Table~\ref{tab:sources}); \\
$^{(3)}$ fractional abundance w.r.t. H$_2$ from $N_{\rm CO}$(H$_2$), given in Table~\ref{tab:sources}; \\
$^{(4)}$ fractional abundance w.r.t. H$_2$ from $N_{\rm Her}$(H$_2$), given in Table~\ref{tab:sources}. \\
\end{table*}
\normalsize

\subsection{Excitation temperatures, FWHM and $V$ of the lines}
\label{temp}

Inspection of Tables~\ref{tab:bestfit-ch3oh},~\ref{tab:bestfit-h2co} and \ref{tab:bestfit-hco} suggests 
that some molecular parameters obtained from \METH, HCO, and \FORM\ are similar, while others are significantly
different. We examine first the best fit \Tex, FWHM and $V$.

\begin{figure}
{\includegraphics[width=12cm]{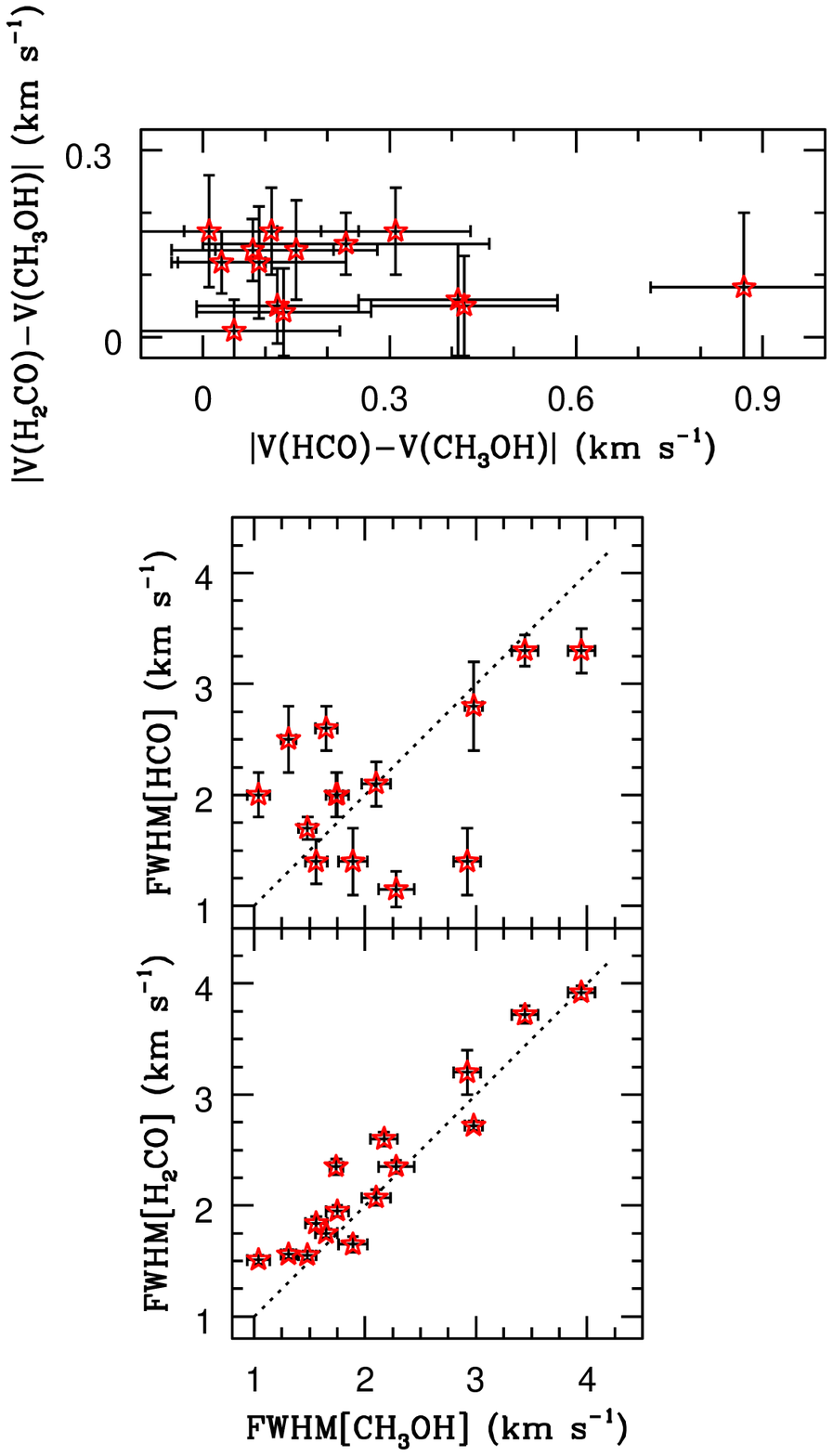}}
      \caption{{\it Top panel}: velocity difference between \FORM\ and \METH\ against that between HCO and 
      \METH. 
      {\it Middle and bottom panels:} comparison between the FWHM measured for HCO and \METH\ (middle) and
      that measured from \METH\ and \FORM\ (bottom). The dotted
      line indicates FWHM[HCO] = FWHM[\METH] or FWHM[\FORM] = FWHM[\METH].
      In all panels, we have not included the HCO data for 19383+2711, for which the resulting fit parameters
      are affected by the unresolved second velocity feature (see Sect.~\ref{hco}).}
\label{fig:velocities}
\end{figure}

The excitation temperatures measured from \FORM\ are larger than those estimated
from \METH, and no correlation is found between the two \Tex\ estimates,
as it can be seen from Fig.~\ref{fig:temperatures}. We have also investigated whether the sources
with higher \Tex, and hence potentially warmer (if \Tex\ is representative of \Tk), are associated 
with gas emitting lines with larger FWHM, and hence more turbulent. As it can be noted from the 
middle and bottom panels in Fig.~\ref{fig:temperatures}, 
we do not find any statistically significant correlation between \Tex\ and FWHM estimated both from 
\FORM\ and \METH.
%Moreover, the \Tex\ estimated from \FORM\
%are more similar to the kinetic temperatures inferred by \citet{bernal21}, while those derived
%from \METH\ are smaller. 
The different \Tex\ may be explained if \METH\ traces gas colder than \FORM. 
On the other hand, the best fit $V$ for \FORM\ and \METH\ are very similar, as they 
differ by less than $\sim 0.2$~\kms\ in all targets (Fig.~\ref{fig:velocities}, top panel),
and also the best fit FWHM are very well correlated (Fig.~\ref{fig:velocities}, bottom panel).
This, hence, suggests that the gas emitting \FORM\ and \METH\ could be made by layers, 
or portions, characterised by different (excitation) temperatures but belonging to a region 
kinematically and spatially coherent. 
%but
%\FORM\ is closer to LTE conditions than \METH, despite the similar
%critical densities of the analysed transitions ($\sim 0.5 -1\times 10^{6}$ \cmc\ for the
%\FORM\ lines). 
%Also, we do not find any statistically significant correlation between \Tex\ and
%FWHM estimated both from \FORM\ and \METH, as it can be noted from the middle and bottom
%panels in Fig.~\ref{fig:temperatures}. This further indicates that the velocity field, or level of
%turbulence, of the gas emitting \FORM\ and \METH\ are not related to the excitation temperature.

The best fit $V$ derived for HCO and \METH\ are different by up to $\sim 0.9$~\kms, i.e.
more than three times the difference in velocity between \METH\ and \FORM\ (Fig.~\ref{fig:velocities},
top panel). Even not including WB89-006, for which the difference is the largest ($\sim 0.9$~\kms), still
the range is $\sim 0.4$~\kms, twice than that found for \FORM\ and \METH.
Also, the lines FWHM derived for HCO do not appear correlated to those of \METH, (Fig.~\ref{fig:velocities},
middle panel), indicating that the observed HCO emission could arise from 
material kinematically, and hence spatially, different from that responsible for the \METH\ and \FORM\ 
emission. In fact, as discussed 
also in \citet{rivilla19}, HCO can be formed on dust grains through hydrogenation of iced CO 
(Tielens \& Hagen~\citeyear{teh82}; Dartois et al.~\citeyear{dartois99}; 
Watanabe \& Kouchi~\citeyear{wek02}; Bacmann \& Faure~\citeyear{bef16}), and in this case be the 
progenitor of \FORM\ and \METH, but also in the gas phase from atomic C and H$_2$O
(Bacmann \& Faure~\citeyear{bef16}; Hickson et al. 2016; Rivilla et al.~\citeyear{rivilla19}). 
In this case, the HCO emission should arise predominantly from an envelope in which C is still 
significantly in atomic form, likely more extended and diffuse than the region from where the emission 
of both \FORM\ and \METH\ arise, which requires most of C to be locked in CO and hence should
be more dense.

Furthermore, as discussed in \citet{peg18}, even in iced HCO, hydrogenation of HCO will form either 
\FORM\ or H$_2$ + CO with equal probability, while, once formed, \FORM\ is fairly robust and can 
quickly form CH$_3$O or CH$_2$OH, from which H addition to finally form \METH\ is fast. For all
these reasons, \FORM\ and \METH\ in cold environments are likely more related than HCO and
\METH, and our observational results are in agreement with this scenario.
\FORM\ can also form in the gas-phase (unlike \METH) in regions where a significant fraction of C 
is not yet locked into CO. Our findings suggest that in these regions most of the C is indeed in form 
of CO, possibly because the present-day O/C ratio is larger than in the Sun neighbourhoods. We will discuss further
this point in Sect.~\ref{distance}.

\subsection{Fractional abundances of \METH}
\label{comparison}
 
From the $N_{\rm CO}{\rm (H_2)}$ listed in Table~\ref{tab:sources}, we estimate the methanol fractional 
abundances with respect to H$_2$, $X$[\METH]. They are listed in Table~\ref{tab:abundances} 
and are in the range 1.1--5.8$\times 10^{-9}$. These values are obtained by smoothing the column 
densities to the largest of the two angular sizes over which they are computed, which
are $\theta_{\rm S}$ for \METH\ (Table~\ref{tab:bestfit-ch3oh}), and 44\asec\ (i.e. the beam size) for 
$N_{\rm CO}{\rm (H_2)}$. 
%Because in some sources $\theta_{\rm S}$ is smaller than 44\asec, and in others is larger, we always
%smoothed to the largest of these two sizes, and hence the derived abundances are values averaged
%over this size.}

We have derived \METH\ abundances also using $N(\rm H_2)$ computed from Herschel Hi-GAL data, 
$N_{\rm Her}\rm (H_2)$, when available (Elia et al.~\citeyear{elia21}). The $N_{\rm Her}(\rm H_2)$ given 
in Hi-GAL, and computed
from the spectral energy distribution of the sources, are averaged within the continuum angular sizes,
$\theta_{\rm c}$, computed from the Herschel $250 \mu$m emission as explained in \citet{elia21}. 
Both $N_{\rm Her}(\rm H_2)$ and $\theta_{\rm c}$ are listed in Table~\ref{tab:sources}, and the 
resulting $X$[\METH] is shown in Table~\ref{tab:abundances}. The two estimates of $X$[\METH] agree 
with each other within a factor 4, even though the computed range in this case is larger (0.6 -- 7.4$\times 10^{-9}$).

We have compared our results with star-forming regions at different metallicities: inner and local Galactic 
targets (representative of Solar and super-Solar metallicities), and OG cores and extragalactic 
sources (representative of sub-Solar metallicities). The list of these sources, with their abundances
and reference works, is given in Table~\ref{tab:abundances}.
Because we cannot constrain robustly the nature and evolutionary stage of our targets yet, we have 
considered a large variety of star-forming regions, from early cold cores embedded in infra-red
dark clouds (IRDCs), to protostellar objects and hot molecular cores in warmer and more
evolved high-mass star-forming regions (HMSFs).
Comparing our sources with targets located in the local and inner Galaxy (Table~\ref{tab:abundances}), 
we do not find significant or systematic differences. The clearest difference is seen towards 
the IRDC cores studied by \citet{vasyunina14}: for these objects the \METH\ abundances are 
on average higher than in our targets. However, our values overlap with the lower edge of their
measured range. Moreover, the HMSF cores observed with angular resolutions similar to
ours (e.g. Minier \& Booth~\citeyear{meb02}, van der Tak et al.~\citeyear{vandertak00}, Gerner 
et al.~\citeyear{gerner14}) show values overlapping with ours. 

Comparing our results to those obtained in other low-metallicity environments, we find a very good
agreement with the OG star-forming cores observed by \citet{bernal21}, as expected since the
two samples have several sources in common.
Interestingly, the \METH\ abundances measured in our study are lower than those measured towards the 
hot core embedded in WB89-789 (1.7$\times 10^{-7}$), as well as in the Small and large Magellanic 
Clouds ($\sim 10^{-8}$). 
This difference, however, is likely due to the higher excitation of the lines used to derive the abundances 
of the mentioned regions, having upper energies typically much larger than 100~K. These
lines are associated with warmer (and likely more compact) gas, enriched in \METH\ upon evaporation of dust
grain mantles, than the one traced by the lines observed in this work, more likely associated with a colder
envelope where a lot of \METH\ is still frozen. 
In fact, \citet{shimonishi18} and \citet{sewilo18} claim that the \METH\ emission arises from hot molecular 
cores embedded inside both the LMC and the SMC. 

Of course, care needs to be taken in these comparisons, given the large number of important assumptions 
(e.g. the assumed gas-to-dust ratio to derive the $N$(H$_2$) column density from the continuum, 
or the CO-H$_2$ conversion factor used to derive $N$(H$_2$) from CO) that can influence significantly 
the abundance estimates (see e.g.~Nakanishi \& Sofue~\citeyear{nes06}; Pineda et al.~\citeyear{pineda13}). 
The probed linear scales are also, in several cases, very different.

\subsection{Fractional abundances of HCO and \FORM\ and their relation with \METH}
\label{abundances-hco-h2co}

\begin{figure}
{\includegraphics[width=8cm]{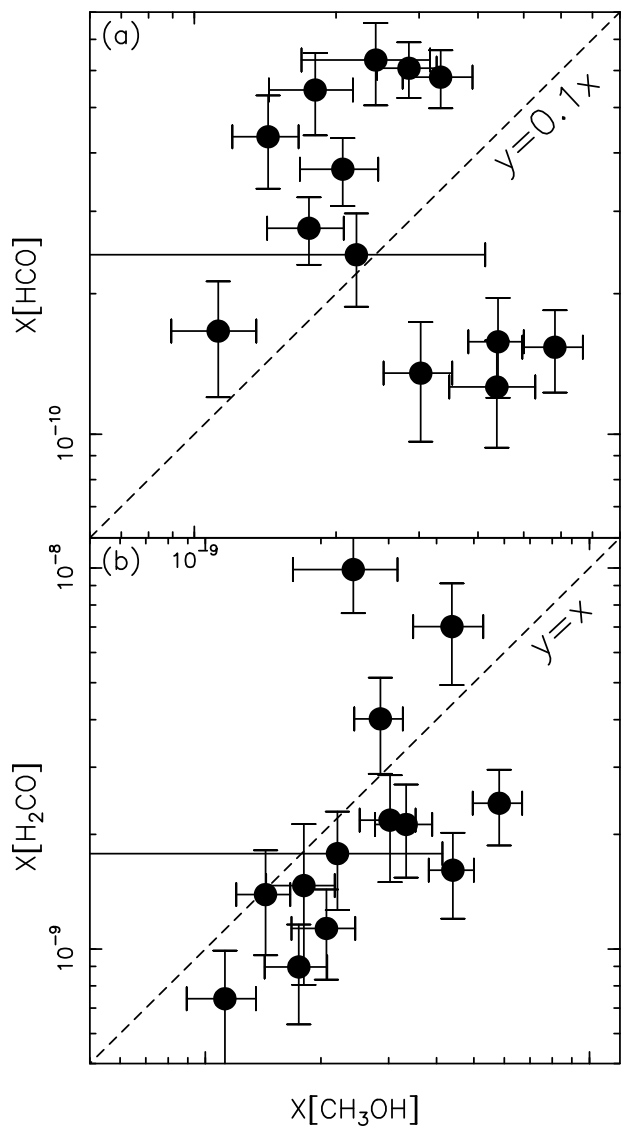}}
      \caption{(a): comparison between the HCO and \METH\ fractional abundances derived from
      $N_{\rm CO}$(H$_2$). The dashed line indicates the locus where y=0.1x;
      {\it (b):} same as panel (a) for \FORM\ and \METH. The dashed line indicates the locus where
      y=x.}
\label{fig:abundances-correlations}
\end{figure}

As for \METH, we derived fractional abundances with respect to H$_2$ for \FORM\ and HCO using
both H$_2$ column density estimates given in Table~\ref{tab:sources}. The results are listed in
Tables~\ref{tab:bestfit-h2co} and~\ref{tab:bestfit-hco}, respectively.
In Fig.~\ref{fig:abundances-correlations} we compare the fractional abundances of methanol with
those of HCO and \FORM. {For $X$[HCO], we use the values derived assuming \Tex\ from \METH, which
are possibly more representative of an extended diffuse envelope
(see Sect.~\ref{hco}), bearing in mind that the estimates using \FORM\ are higher by a factor $\sim 1.5 -3$.}
The plot indicates that $X$[HCO] and $X$[\METH] are not correlated. We find a tentative 
positive correlation between $X$[H$_2$CO] and $X$[\METH] (Pearson's $\rho$ correlation coefficient 
$\sim 0.3$). 
This would support our previous claim (Sect.~\ref{temp}), based on the $V$ and FWHM of the lines, 
that the \METH\ emission is more likely related to \FORM\ emission than to HCO emission. 
However, care needs to be taken in the interpretation of these plots since the correlation is tentative, 
it is strongly influenced by one single source, WB89-399,
and \FORM\ is also known to form in the gas phase (unlike \METH) from regions rich in hydrocarbons, 
i.e. where C is not yet completely locked in CO (see e.g. Chacon-Tanarro et al.~\citeyear{chacon19}). 

\subsection{Abundance variations with the Galactocentric distance}
\label{distance}

In Fig.~\ref{fig:comparison} we show the \METH, HCO, and \FORM\ abundances against the source 
Galactocentric distances, $R_{\rm GC}$. The filled symbols show the abundances calculated using 
$N_{\rm CO}$(H$_2$), while the empty ones illustrate those calculated using $N_{\rm Her}$(H$_2$) 
(Table~\ref{tab:sources}). 
Again, for HCO we show the results obtained using as \Tex\ the excitation temperatures of \METH,
being more likely representative of an envelope more extended than that traced by \FORM.
A simple linear regression fit to the data (solid line) shows an almost flat line, indicating that all 
abundances seem independent of $R_{\rm GC}$. This suggests that the decreasing metallicity 
towards the external part of the Galaxy should not have effect on the abundance of \METH. 
This result was already found by \citet{bernal21}. 
The main novelty of this study is that also for HCO and \FORM, two progenitors of \METH, 
the abundance does not decrease at metallicities lower than the Solar one. 

However, both abundance estimates are based on crucial assumptions: 
in the method of \citet{blair08}, a CO-H$_2$ conversion factor is assumed independent on
$R_{\rm GC}$, while departures from a constant value are both observed and expected in both
the Milky Way and external galaxies (Bolatto et al.~\citeyear{bolatto13}; Pineda et al.~\citeyear{pineda13}; 
Casasola et al.~\citeyear{casasola17};~\citeyear{casasola20}); in the method of \citet{elia21}, 
a constant gas-to-dust ratio of 100 is assumed, which is also expected to change radially
(e.g.~Magrini et al.~\citeyear{magrini11}). For the former, a Galactocentric trend has 
been proposed by Nakanishi \& Sofue~(\citeyear{nes06}, see their Eq.(2)), from which the 
CO-H$_2$ conversion factor would vary by a factor $\sim 2$ from 13 to 19~kpc, still consistent
with an almost flat trend of the molecular abundances with $R_{\rm GC}$.
For the latter, we have investigated how the abundances change considering the Galactocentric 
increasing trend found by \citet{giannetti17} for the gas-to-dust ratio. In Fig.~\ref{fig:comparison}, 
we plot the abundances of \METH, HCO, and \FORM\ derived from $N_{\rm Her}\rm (H_2)$
corrected according to Eq. (2) in \citet{giannetti17}:
\begin{equation}
\label{eq:giannetti}
Log(\gamma) = (0.087 \pm 0.007)\times R_{\rm GC} + (1.44 \pm 0.03) \;,
\end{equation}
where $\gamma$ is the gas-to-dust ratio. We do not include the systematic uncertainties for simplicity.
Please note that Eq.~(\ref{eq:giannetti}) assumes the same Solar Galactocentric radius as we do.

As expected, now all molecules show abundances decreasing with $R_{\rm GC}$. However,
this overall decrease does not seem to imply a reduced efficiency in the formation of these organics, as
we will discuss in the following sub-sections.

\subsubsection{\METH}

Let us start examining the case of \METH.
A linear regression fit applied to the points plotted in panel (a) of Fig.~\ref{fig:comparison} gives:
\begin{equation}
\label{eq:fit-ch3oh}
X[{\rm CH_3OH}] = (-1.14\times 10^{-10})\times R_{\rm GC} + 2.45\times 10^{-9} \;,
\end{equation}
\noindent
which implies that $X$[\METH] decreases by a factor $\sim 5$ from 8.34~kpc (the Sun's Galactocentric distance) 
to 19~kpc, extrapolating the trend found in the OG to the local Galaxy. According to the gradients
given in \citet{arellano20}, the [O/H] and [C/H] ratios at 19~kpc are $\sim 1.2\times 10^{-4}$ and $\sim 4.7\times 10^{-5}$, 
i.e. $\sim 3$ and $\sim 6$ times lower, respectively, than the values at the Solar circle 
([O/H]$\sim 3\times 10^{-4}$ and [C/H]$\sim 2.8\times 10^{-4}$). 
The more recent gradients given in M\'endez-Delgado et al.~(\citeyear{mendez22}) provide [O/H]$\sim 1.1\times 10^{-4}$
and [C/H]$\sim 3.9\times 10^{-5}$, respectively, at 19~kpc, i.e. $\sim 3$ and $\sim 7$ times lower than Solar 
([O/H]$\sim 3.1\times 10^{-4}$ and [C/H]$\sim 2.6\times 10^{-4}$).

Therefore, the observed scaling factor of $X$[\METH] when applying the gas-to-dust ratio correction is 
consistent with that of the [C/H] ratio, or even marginally smaller than it. This suggests that
the "efficiency" in the formation of \METH, scaling with the availability of the parent element C, is
at least as high as in the local Galaxy.

\subsubsection{HCO and \FORM}

Similarly, for $X$[HCO] and $X$[H$_2$CO] we find a negligible decrease with $R_{\rm GC}$
without considering the Galactocentric variation of the gas-to-dust ratio, and a decrease similar to that
of \METH\ when applying it. In fact, the linear regression fits to the points plotted in Fig.~\ref{fig:comparison} 
provide:
\begin{equation}
\label{eq:fit-hco}
X[{\rm HCO}] = (-1.38\times 10^{-11})\times R_{\rm GC} + 2.95\times 10^{-10} \;;
\end{equation}

\begin{equation}
\label{eq:fit-h2co}
X[{\rm H_2CO}] = (-1.18\times 10^{-10})\times R_{\rm GC} + 2.51\times 10^{-9} \;.
\end{equation}

These relations imply that both $X$[HCO] and $X$[H$_2$CO] decrease by a factor $\sim 5.5$  
from 8.34~kpc to 19~kpc. However, several caveats must
be taken into consideration. First, the Galactocentric trend for the gas-to-dust ratio is associated 
with large systematic and statistical uncertainties (Giannetti et al.~\citeyear{giannetti17}), 
and local values may deviate significantly from the proposed trend. Second, our linear regression 
fit is obtained in the range $R_{\rm GC}\sim 13 - 19$~kpc and could not be appropriate 
down to $R_{\rm GC}\sim 8.34$~kpc. Third, the comparison with Galactocentric gradients needs
to be taken with caution because the elemental abundances are still under debate. For example,
the already mentioned [C/H] decrease by a factor 6 from the Solar circle to 19~kpc
found by Arellano-C\'ordova et al.~(\citeyear{arellano20}) is consistent with the 5.5 scaling factor 
predicted by Eqs.~(\ref{eq:fit-hco}) and (\ref{eq:fit-h2co}).
However, the more recent gradients derived by \citet{mendez22} indicate a scaling factor of 
$\sim 7$ for [C/H], that is marginally higher than our estimates. 
%On the other hand, oxygen is expected to decrease only by
%a factor $\sim 3$, due to the fact that C and O have different origins, 
%with contributions to their 
%production by stars with different masses, and thus lifetimes (see, e.g. Randich \& Magrini~\citeyear{rem21} 
%for a review). Therefore, it is natural to expect that they do not follow the same gradient.}

Therefore, notwithstanding the above mentioned uncertainties, one can conclude that the formation of these 
molecules is not inhibited in low-metallicity regimes, because all of them show abundances that are in 
line with the initial local abundances of the parent elements, and very likely not smaller than them.
As discussed in \citet{blair08},
\citet{bernal21}, and in paper I, this finding, and the additional evidence that Earth-like
planets are ubiquitously found in the Galaxy (Sect.~\ref{intro}), suggest that an appropriate "ground" 
for the formation of habitable planets is present also in the OG, and calls for a redefinition of the
GHZ that takes requirements other than metallicity into account.

\subsubsection{Abundance ratios}

\begin{figure}
{\includegraphics[width=8cm]{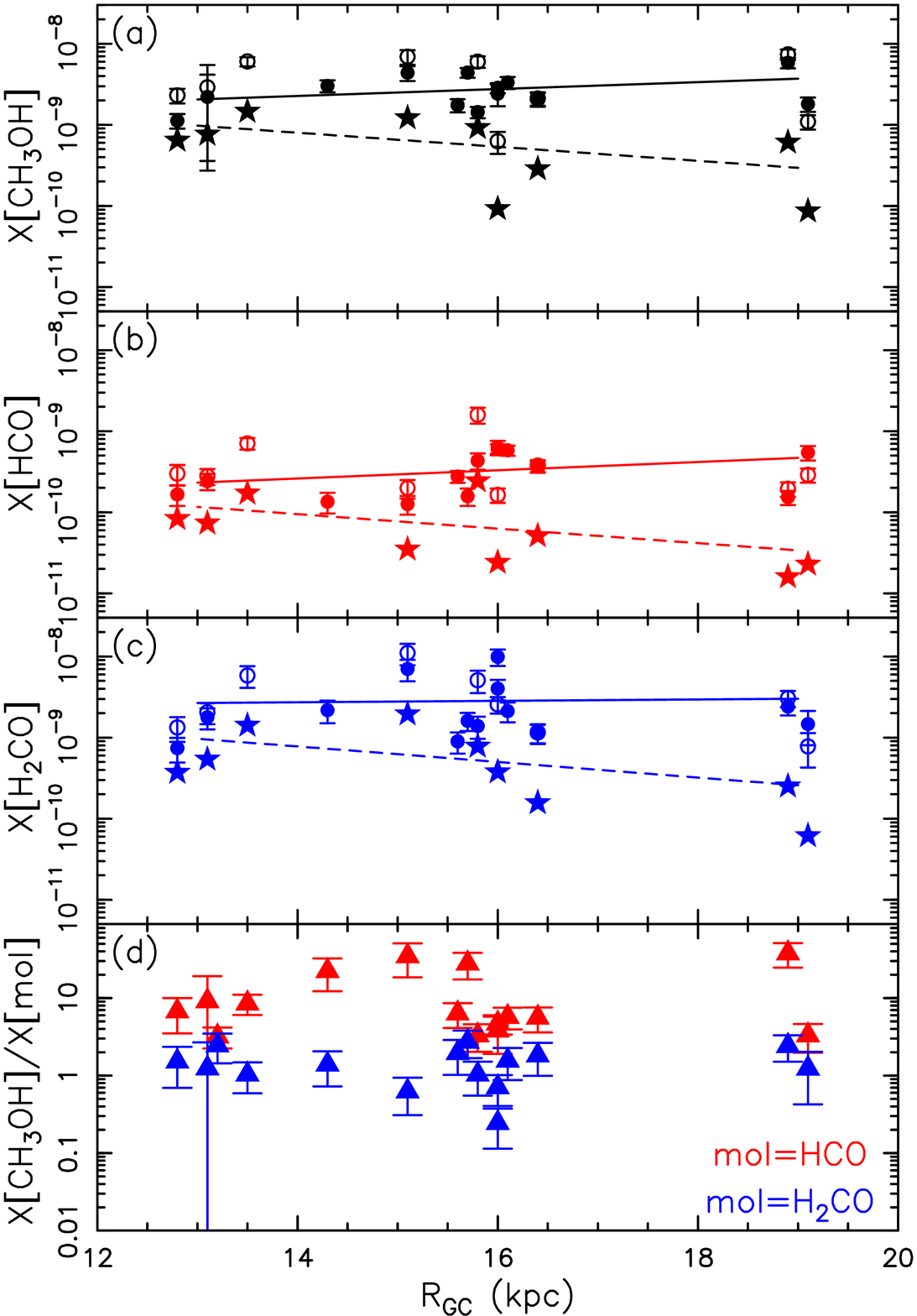}}
      \caption{{\it Panels (a), (b) and (c)}: fractional abundances of, from top to bottom, \METH, HCO, 
      and \FORM, as a 
      function of $R_{\rm GC}$. The fractional abundances are derived from $N_{\rm CO}$(H$_2$) given
      in \citet{blair08} (filled circles, Table~\ref{tab:sources}) and from $N_{\rm Her}$(H$_2$) given
      in \citet{elia21} (empty circles, Table~\ref{tab:abundances}). The stars correspond to the values 
      computed from $N_{\rm Her}$(H$_2$) corrected for the Galactocentric trend for the gas-to-dust ratio 
      given in \citet{giannetti17}. In all panels, the solid and dashed lines connect the values obtained at
      13 and 19~kpc from a linear regression fit
       applied to the points computed from $N_{\rm Her}$(H$_2$) without and with corrections, respectively,
      for the Galactocentric trend of the gas-to-dust ratio.
      \newline
      {\it Panel (d)}: abundance ratios $X$[CH$_3$OH]/$X$[HCO] (red triangles) and 
      $X$[CH$_3$OH]/$X$[H$_2$CO] (blue triangles) as a function of $R_{\rm GC}$.}
\label{fig:comparison}
\end{figure}

Finally, the column density ratios $N$[CH$_3$OH]/$N$[HCO] as a function of $R_{\rm GC}$
are shown in the bottom panel of Fig.~\ref{fig:comparison}, and range from $\sim 3$ (19383+2711) to 
$\sim 38$ (WB89-621), with an average value of $\sim 12$ (median $\sim 6.4$). The ratios 
$N$[CH$_3$OH]/$N$[H$_2$CO], shown in the same plot are in between $\sim 0.2$ (WB89-399) and 
$\sim 2.7$ (WB89-437), with an average value of $\sim 1.5$ (median 1.4). The latter values are consistent 
with those measured by \citet{bernal21}. Both ratios do not change with $R_{\rm GC}$, indicating
once more that the chemistry that connects these species do not seem to vary (at least in an obvious
way) with the distance from the Galactic Centre.

\subsection{Methanol formation at low metallicity}
\label{formation}

The formation of \METH\ in dense star-forming cores is usually attributed to be mostly
due to hydrogenation of CO which, on the surfaces of dust grains, forms sequentially
HCO, \FORM, CH$_2$OH, and finally \METH. This is largely believed to be the most
important formation route, given the inefficiency of gas phase routes at low temperatures
(e.g. Garrod et al.~\citeyear{garrod06}). If the main formation route is the same in the
low metallicity environment of the OG, the large abundances measured in this study
would suggest an active surface chemistry, followed by desorption mechanisms. In
this respect, mantle evaporation either from internal protostellar activity, or from 
external processes, are typically invoked. Reactive desorption can also be important.
As described by Vasyunin et al. (~\citeyear{vasyunin17}), reactive desorption becomes 
very efficient in regions where icy mantles become CO-rich (i.e. in dense gas where the 
catastrophyc CO freeze-out takes place and CO ice becomes the dominant component of the icy mantles).   
In paper I, we have reported the detection of SiO $J=2-1$ in 46$\%$ of the 35 targets.
Among the 15 sources studied in this paper, 10 are clearly or tentatively associated with 
SiO emission, while five (WB89-789, WB89-006, WB89-076, WB89-080, and WB89-283) 
are not. On the other hand, all sources but 19383+2711 are associated with high-velocity
wings in the \HCOp\ $J=1-0$ line (paper I). This suggests that, even in the OG, the origin 
of \METH\ emission could be connected to evaporation of grain mantles caused by 
protostellar outflows. The non detection of SiO in the five targets mentioned above may 
be due to insufficient sensitivity, given that these objects are not the most intense of the 
sample in the molecular lines analysed both in this paper 
(see Figs.~\ref{fig:spectra-Fig1} -- \ref{fig:spectra-H2CO-Fig2}) and in paper I. 
However, first, the methanol line widths for all sources are narrow with respect to what is expected from
material freshly released from outflows. Second, the full analysis of both SiO and \HCOp\ 
lines goes beyond the scope of this work and will be performed in a forthcoming paper 
(Fontani et al. in prep.). Moreover, this result could also be influenced by the source selection
we have performed in this study, and should be corroborated with a higher statistics. 

Chemical models with modified (lower) metallicities were developed in the past to
interpret the formation of methanol and other COMs in the Magellanic Clouds as
representatives of low-metallicity environments. \citet{aeh15} modelled dense and cold
clouds of the LMC. They found that some observed results, especially for methanol, 
are better matched if these regions currently have lower temperatures. This is in
agreement with the low temperature associated with our observed methanol emission.
Moreover, \citet{aeh16} modelled dense clouds of the SMC, in which the metallicity
is lower than the Solar one by a factor 5, and found that for species produced fully 
(e.g., \METH) or partially on the grain surfaces (e.g., \FORM), the predicted abundances 
are not just metallicity-scaled values but they change in a more complex way. 
Our observations also indicate that care needs to be taken in the comparison with
scaled-metallicity values, because different assumptions in the derivation of the
abundances can bring to different conclusions (see Sect.~\ref{distance}).

\citet{peg18} used the gas-grain chemical model MAGICKAL to model the chemistry in
clouds of the Magellanic clouds where massive young stellar objects are forming.
They discuss the formation and evolution of several species chemically connected
to \METH, and of \METH\ itself, and conclude that the methanol abundance is even
found to be enhanced in low-metallicity environments. In fact, their models predict
that the amount of \METH\ with respect to CO increases as the elemental C
decreases, thus indicating a more efficient abundance at lower metallicities. This effect
could be due to the smaller C/O ratio (see Sect.~\ref{distance}), so that the bulk of C 
is in the form of CO, required to form \METH. If confirmed, this would imply a lower
abundance of carbon-chains, or other carbon rich molecules. The exploitation of the
detected carbon-rich species, listed in paper I, among which {\it c-}C$_3$H$_2$,
C$_4$H, and CCS, will be performed in forthcoming papers (Fontani et al., in prep.).

However, when modelling the chemistry, several ingredients need to be considered and
many crucial ones could vary significantly with the Galactocentric radius. For example,
as also discussed, e.g., in ~\citet{viti20}, the abundances of molecules containing carbon 
in external galaxies known to be associated with different visual extinctions, cosmic-ray 
ionisation rates, and/or Ultraviolet (UV) radiation fields are predicted to change by orders of 
magnitude. Because the parameters aforementioned are expected to vary within the Milky Way 
as well (e.g. both the interstellar UV field and the cosmic-ray
 ionisation rate are expected to be lower in the OG due to less numerous of massive
stars and of supernovae explosions), to appropriately model the chemistry a thorough analysis 
of the variations with $R_{\rm GC}$ of all these key parameters is absolutely needed.
This discussion is particularly urgent, since important complex species in external galaxies are
now detected easily thanks to the available powerful (sub-)mm telescopes (see e.g. the first results 
of the ALCHEMI ALMA large program, Mart\'in et al.~\citeyear{martin21}), 
showing that a rich chemistry can develop even in external galaxies.
This goes beyond the scope of this paper, but strongly calls for both observational works
to measure and constrain these parameters as a function of the Galactic radius, and 
theoretical works aimed at identifying the physical conditions that mostly affect the chemistry
in such modified environments. In particular, chemical evolution models (e.g.~Romano et 
al.~\citeyear{romano20}) could be used in the future to estimate the amount and density 
of massive stars, important sources of both UV photons and cosmic rays, as a function of 
the Galactocentric radius.

\section{Conclusions}
\label{conc}

\begin{table}
\label{tab:abundances}
\setlength{\tabcolsep}{1.5pt}
\begin{center}
\caption[]{Abundances of \METH\ calculated in this work, and towards other star-forming regions 
in the Galaxy, as well as in external galaxies.}
\begin{tabular}{ccc}
\hline \hline
%   & \multicolumn{2}{c}{$X$(\METH) - this work}  & &  \\
     & $X_{\rm CO}$[\METH]$^{(a)}$   & $X_{\rm Her}$[\METH]$^{(b)}$   \\
     &   $\times 10^{-9}$                   &       $\times 10^{-9}$                     \\
\hline
WB89-379 & 2.1(0.4) & 2.1(0.4) \\
WB89-380 & 2.9(0.4) & --       \\
WB89-391 & 3.3(0.6)  & --      \\
WB89-399 & 2.4(0.7) & 0.6(0.2) \\
WB89-437 & 4.4(0.6) & --       \\
WB89-501 & 1.8(0.3) & --       \\
WB89-621 & 5.8(0.9) & 7.4(1.1) \\
WB89-789 & 1.8(0.4) & 1.1(0.2) \\
19383+2711 & -- & --           \\
19423+2541 & -- & 6.0(0.7)     \\
WB89-006 & 3.0(0.5) & --       \\
WB89-035 & 2.2(1.3) & 2.9(2.5) \\
WB89-076 & 4.4(0.9) & 6.9(1.4) \\
WB89-080 & 1.1(0.2) & 2.3(0.5) \\
WB89-283 & 1.4(0.2) & 6.0(1.0) \\
\hline
  Inner and local Galaxy & $X$[\METH] &  Ref.$^{(c)}$ \\
  \hline
 IRDC cores &  0.52--65$\times 10^{-9}$  & (1) \\
 IRDC cores &  $\leq 1 \times 10^{-9}$ &  (2) \\
 HMSF cores &  0.07--1.5$\times 10^{-9}$ & (3) \\
 HMSF cores &  0.9$\times 10^{-9}$ &  (2) \\
 Hot cores     &  2.6$\times 10^{-8}$ & (2) \\
 HMSF cores &  0.4--24$\times 10^{-9}$ & (4) \\
\hline
  OG cores and  & & \\
  low-metallicity galaxies &  & \\ 
\hline
OG cores &  0.2-4.9$\times 10^{-9}$ & (5) \\
  WB89-789 hot core & 1.7$\times 10^{-7}$ & (6) \\
  Small Magellanic Cloud &  0.5--1.5$\times 10^{-8}$ & (7) \\ 
  Large Magellanic Cloud &  2$\times 10^{-10}$--5.6$\times 10^{-8}$ & (8, 9) \\
\hline
\end{tabular}
\end{center}
$^{(a)}$ from $N_{\rm CO}$(H$_2$), given in Table~\ref{tab:sources}; \\
$^{(b)}$ from $N_{\rm Her}$(H$_2$), given in Table~\ref{tab:sources}; \\
$^{(c)}$ References: (1) Vasyunina et al.~(\citeyear{vasyunina14}, averaged on an angular scale of 29\asec); (2) Gerner et al.~(\citeyear{gerner14}, averaged on an angular scale of 11\asec); (3) Minier \& Booth~(\citeyear{meb02}, averaged on an angular scale of 34\asec); (4) van der Tak et al.~(\citeyear{vandertak00}, averaged on an angular scale of 18\asec); (5) Bernal et al.~(\citeyear{bernal21}), averaged on a beam of 63\asec; (6) Shimonishi et al.~(\citeyear{shimonishi21}, averaged on an angular scale of 1.9\asec); (7) Shimonishi et al.~(\citeyear{shimonishi18}, averaged on an angular scale of $\sim 0.6$\asec); (8) Sewi\l{}o et al.~(\citeyear{sewilo18}, averaged on an angular scale of $\sim 0.8$\asec); Sewi\l{}o et al.~(\citeyear{sewilo22}, averaged on an angular scale of $\sim 0.8$\asec). \\
\end{table}
\normalsize

We detected \METH, HCO, and \FORM\ emission associated with 15 star-forming regions of the 
outer Galaxy. Derived angular diameters, excitation temperatures and line widths of \METH\ indicate 
that the emission is dominated by a cold and quiescent gaseous envelope. The \METH\ fractional abundances 
are in the range $\sim 0.6 - 7.4\times 10^{-9}$. These values are consistent with similar star-forming
regions in the local and inner Galaxy. 
We find that some \METH\ line parameters, such as centroid velocity, FWHM, and (less obviously) fractional
abundance with respect to H$_2$, are correlated to those of \FORM, while they do not appear correlated 
to those of HCO. This may indicate that hydrogenation of \METH\ from iced \FORM, followed by 
evaporation or other desorption mechanisms from grain mantles, can be the relevant source of 
\METH\ also in the OG. On the other hand, the HCO emission we observe can also be significantly produced in 
the gas phase from routes not involving \FORM\ and \METH. However, in these observations 
the \METH\ emission is clearly associated with an extended and relatively quiescent envelope 
rather than with shocked or sputtered material, hence this conclusion needs to be corroborated 
by higher-angular resolution observations. The total column densities and fractional abundances
with respect to H$_2$ indicate that the production of \METH\ is not inhibited even at Galactocentric
distances of $\sim 19$~kpc, where the carbon abundance is estimated to be lower by a factor $\sim 6-7$
with respect to the Solar one. In fact, even considering corrections to the estimated abundances
taking the variations of the gas-to-dust ratio with $R_{\rm GC}$ into account, our abundances are
in line with metallicity-scaled values. The high abundance of \METH\ at large $R_{\rm GC}$ could
be due to the smaller C/O ratio at these large Galactocentric distances, that implies that the bulk of C 
is in the form of CO, required to form \METH.
Our results confirm that organic chemistry is active and efficient even in the outermost star-forming  
regions of the Milky Way, and support the idea that the outer boundaries of the Galactic Habitable Zone 
need to be re-discussed in light of the capacity of the interstellar medium to form organic molecules even 
at such low metallicities.

\begin{acknowledgements}
We thank the anonymous Referee for their valuable and constructive comments.
F.F. is grateful to the IRAM 30m staff for their precious help during the observations.
L.C. has received partial support from the Spanish State Research Agency (AEI; project number 
PID2019-105552RB-C41). V.M.R. acknowledges support from the Comunidad de Madrid through 
the Atracci\'on de Talento Investigador Modalidad 1 (Doctores con experiencia) Grant 
(COOL: Cosmic Origins of Life; 2019-T1/TIC-15379).
This publication was supported by the European Union Horizon 2020
research and innovation programme under grant agreement No 730562 and
grant agreement No 101004719 [ORP] (RadioNet). 
\end{acknowledgements}

{}

\newpage 

\onecolumn
\begin{appendix}
\section{Spectra}
%\section{Tables}
\label{appa}

We show in this appendix the spectra of \METH, HCO and H$_2$CO analysed in this work (see Sect.~\ref{sample}).

\FloatBarrier
\begin{figure*}[h]
{\includegraphics[width=15cm]{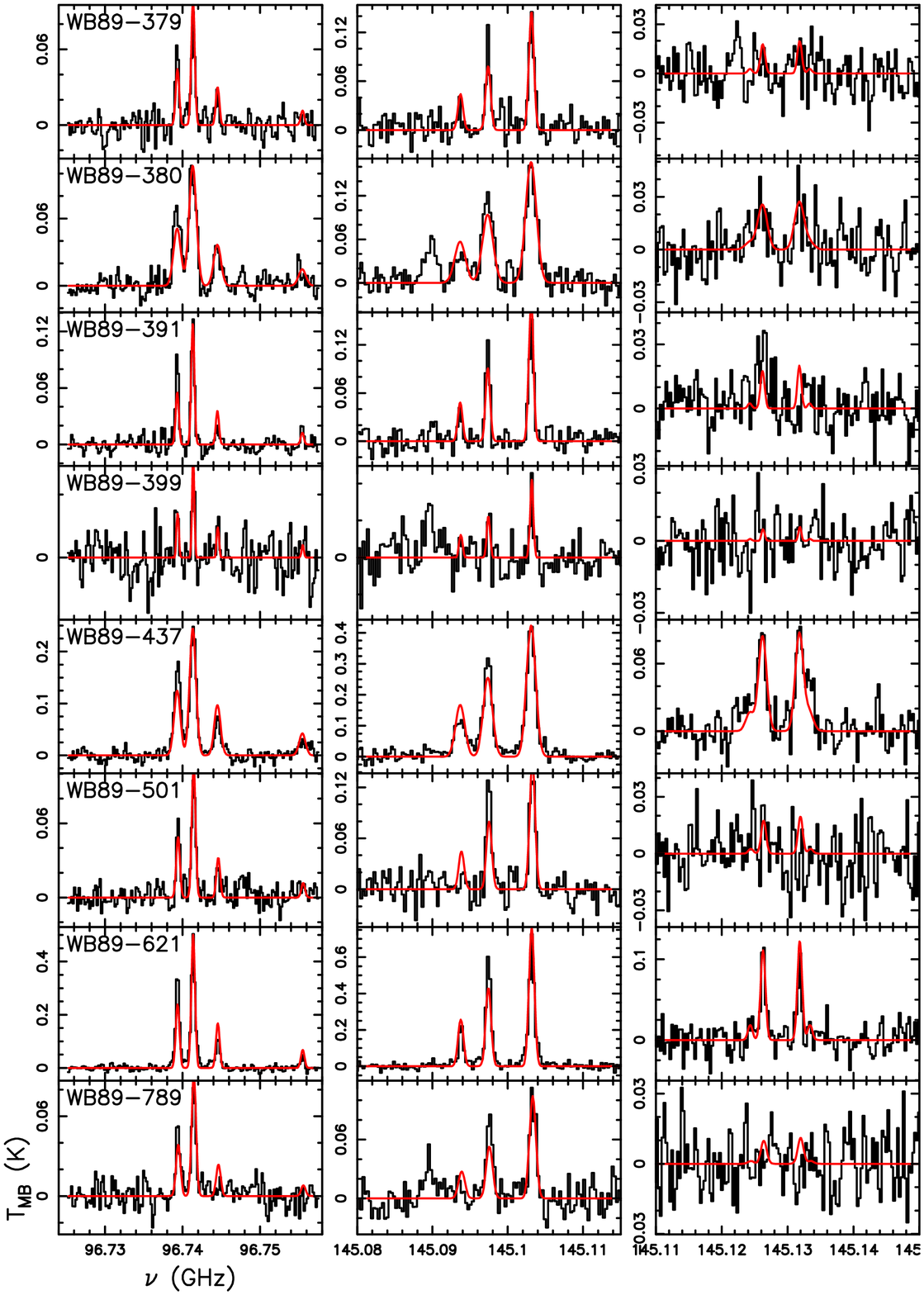}}
      \caption{Spectra of \METH\ lines identified in the 3 and 2~mm bands (Table~\ref{tab:lines}) of the IRAM-30m 
      telescope
      towards the eight first sources listed in Table~\ref{tab:sources}. The red curve in each frame represents the 
      best fit to the lines performed with {\sc madcuba} (see Sect.~\ref{analysis}).}
\label{fig:spectra-Fig1}
\end{figure*}

\FloatBarrier
\begin{figure*}
{\includegraphics[width=15cm]{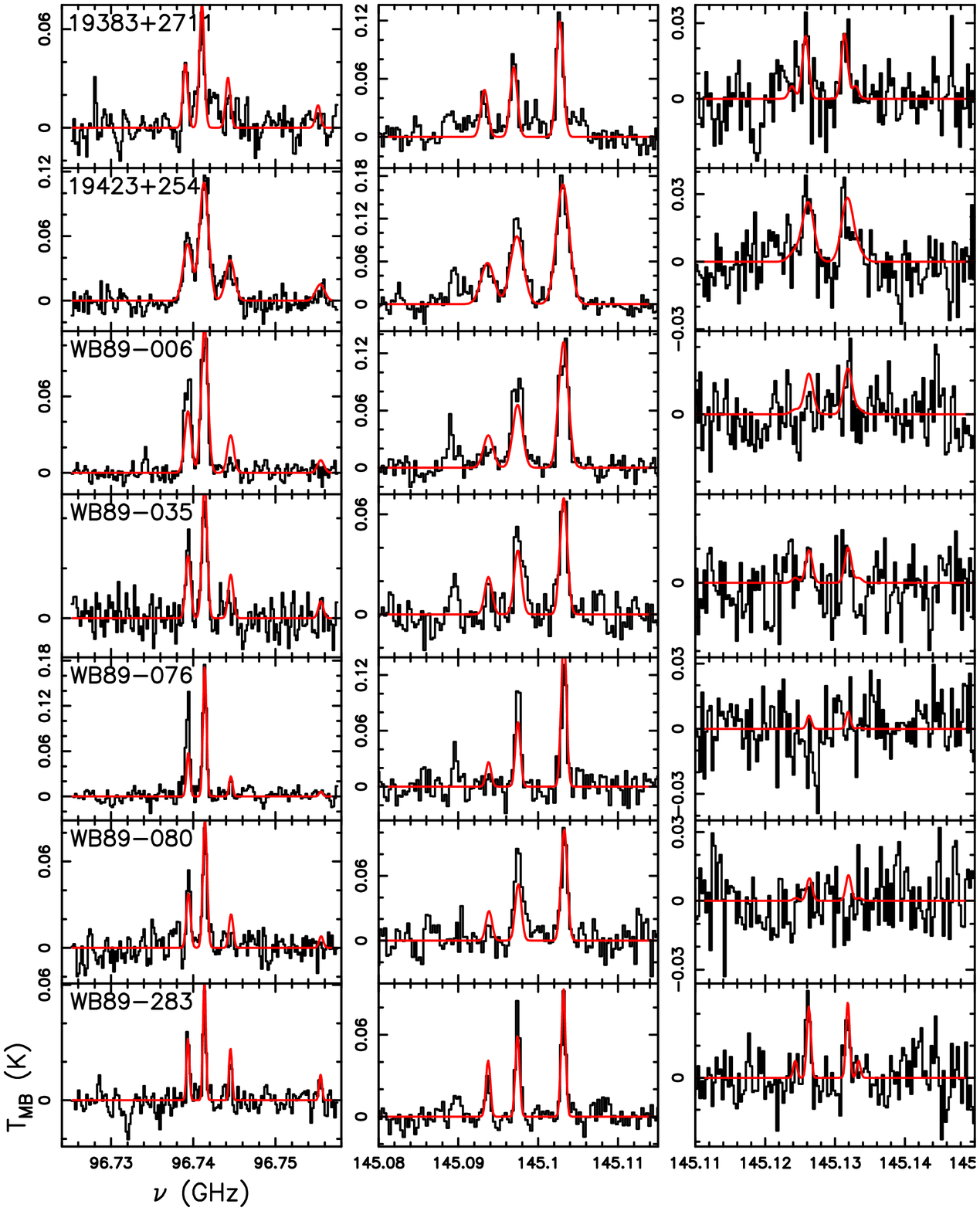}}
      \caption{Same as Fig~\ref{fig:spectra-Fig1} for the remaining seven sources.}
\label{fig:spectra-Fig2}
\end{figure*}

\FloatBarrier
\begin{figure*}
{\includegraphics[width=15.5cm]{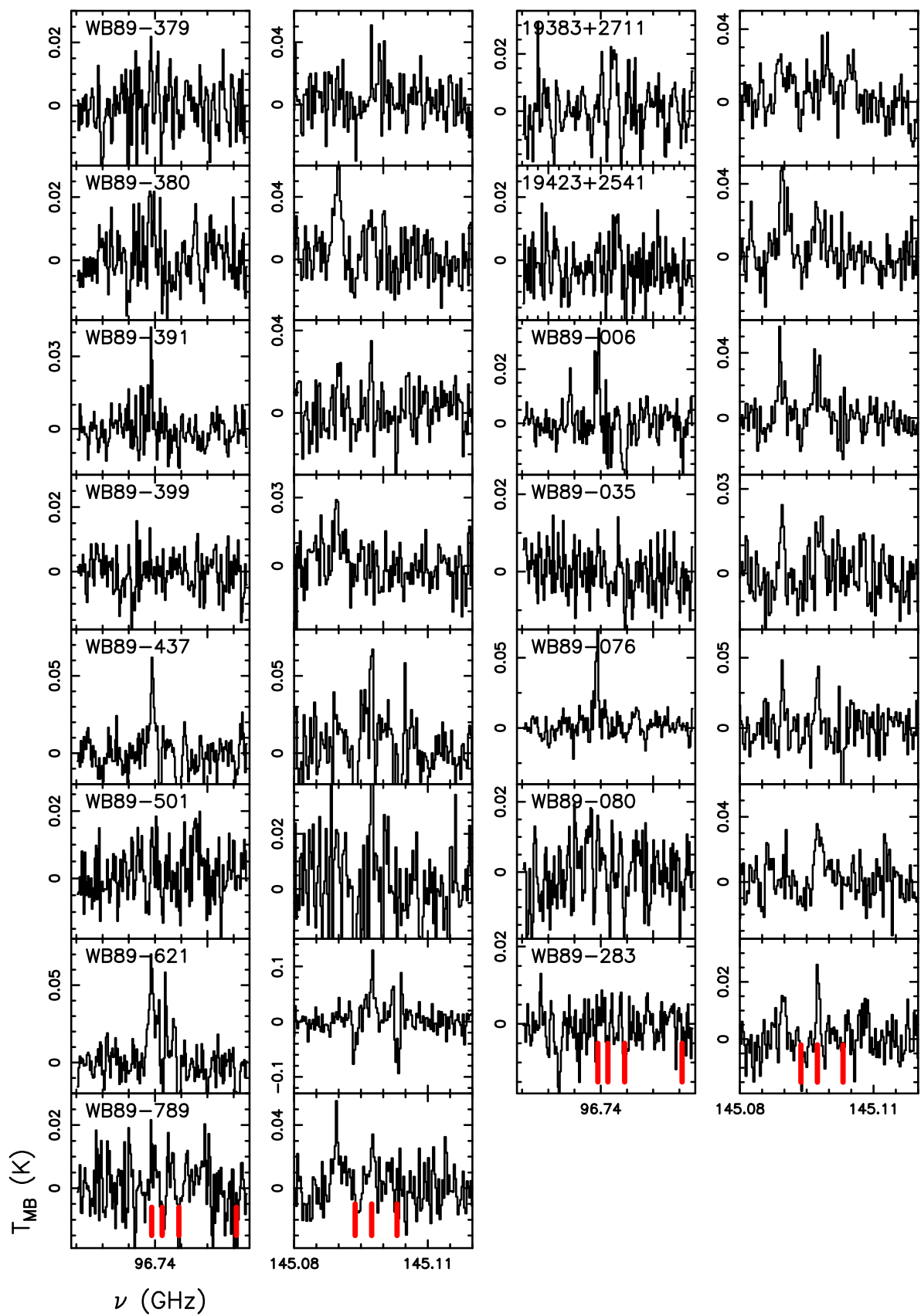}}
      \caption{Residuals obtained from the best fits shown in Figs.~\ref{fig:spectra-Fig1} and \ref{fig:spectra-Fig2}.
      At 2~mm, we only show the residuals of the lines shown in the second panel
      of Figs~\ref{fig:spectra-Fig1} and \ref{fig:spectra-Fig2}. The frequency of the fitted lines are indicated
      by vertical red lines.}
\label{fig:residuals}
\end{figure*}

\FloatBarrier
\begin{figure*}
{\includegraphics[width=15cm]{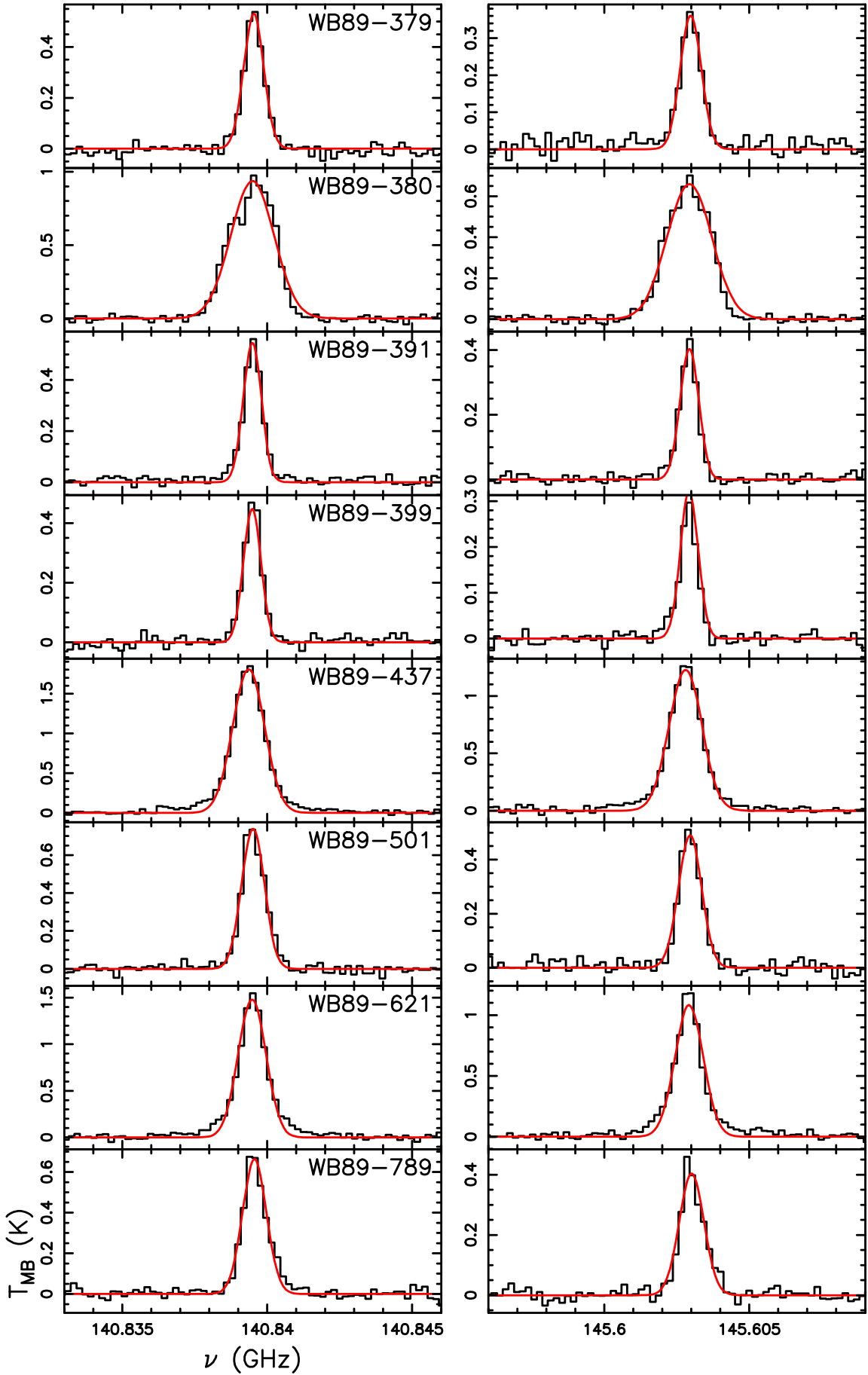}}
      \caption{Spectra of the H$_2$CO lines listed in Table~\ref{tab:lines} observed at 2~mm with the IRAM-30m
      telescope  
      towards the eight first sources listed in Table~\ref{tab:sources}. The red curve in each frame represents the 
      best fit to the lines performed with {\sc madcuba} (see Sect.~\ref{analysis}).}
\label{fig:spectra-H2CO-Fig1}
\end{figure*}

\FloatBarrier
\begin{figure*}
{\includegraphics[width=15cm]{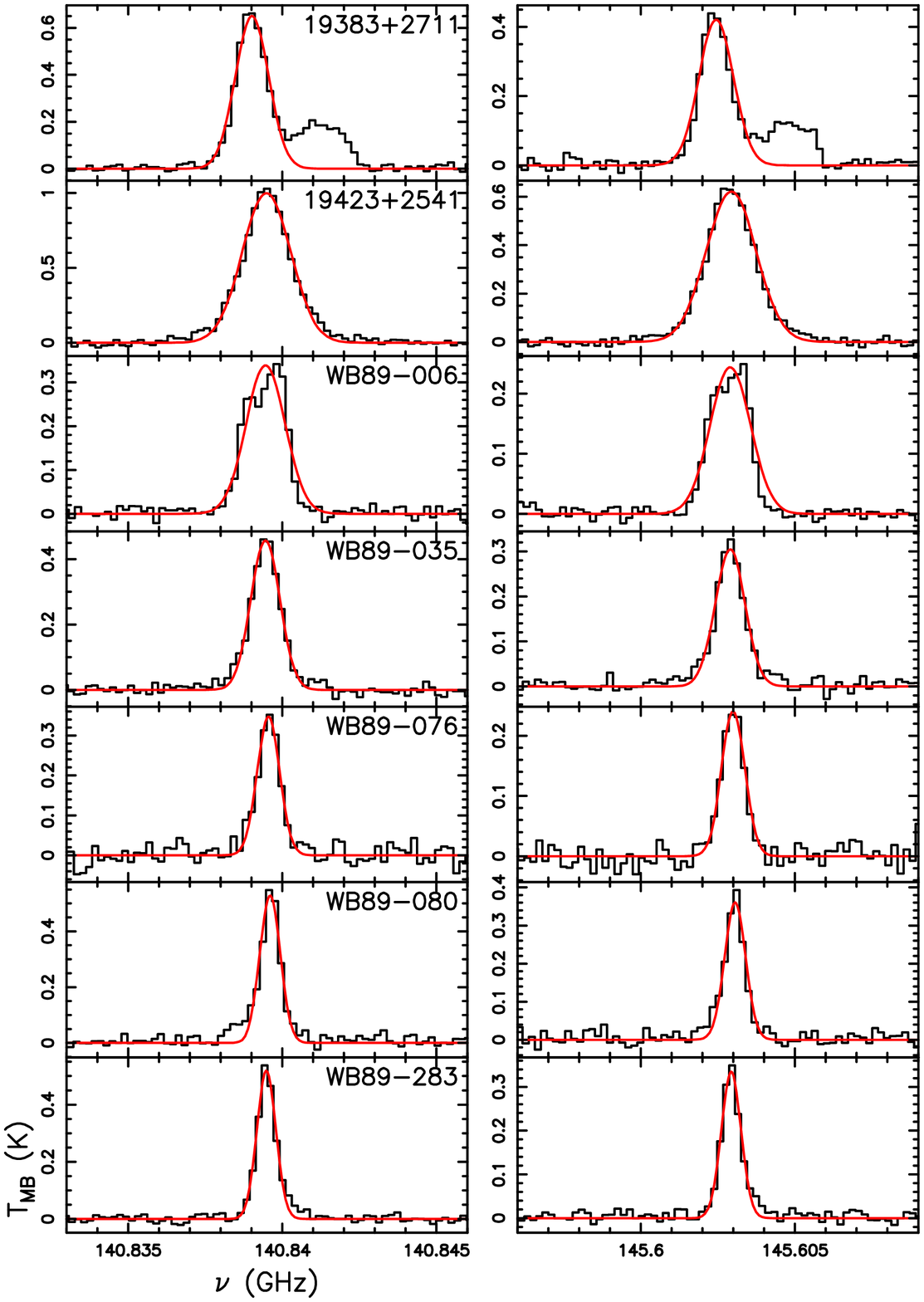}}
      \caption{Same as Fig~\ref{fig:spectra-H2CO-Fig1} for the remaining seven sources.}
\label{fig:spectra-H2CO-Fig2}
\end{figure*}

\FloatBarrier
\begin{figure*}
{\includegraphics[width=15cm]{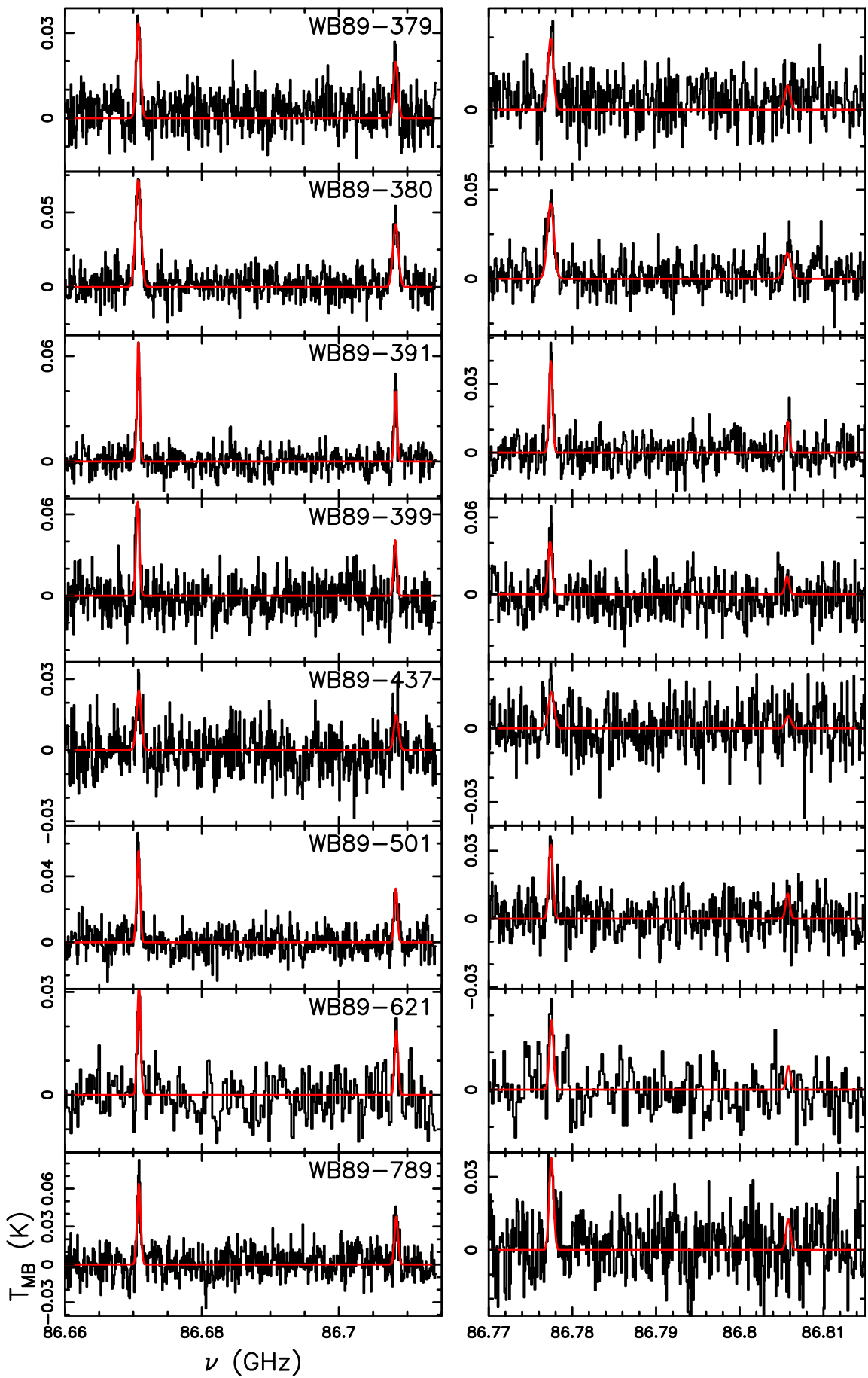}}
      \caption{Spectra of the HCO lines listed in Table~\ref{tab:lines} observed at 3~mm with the IRAM-30m
      telescope  
      towards the eight first sources listed in Table~\ref{tab:sources}. The red curve in each frame represents the 
      best fit to the lines performed with {\sc madcuba} by fixing \Tex\ at the excitation temperature of \METH\
      (see Sects.~\ref{analysis} and \ref{hco}).}
\label{fig:spectra-HCO-Fig1}
\end{figure*}

\FloatBarrier
\begin{figure*}
{\includegraphics[width=15cm]{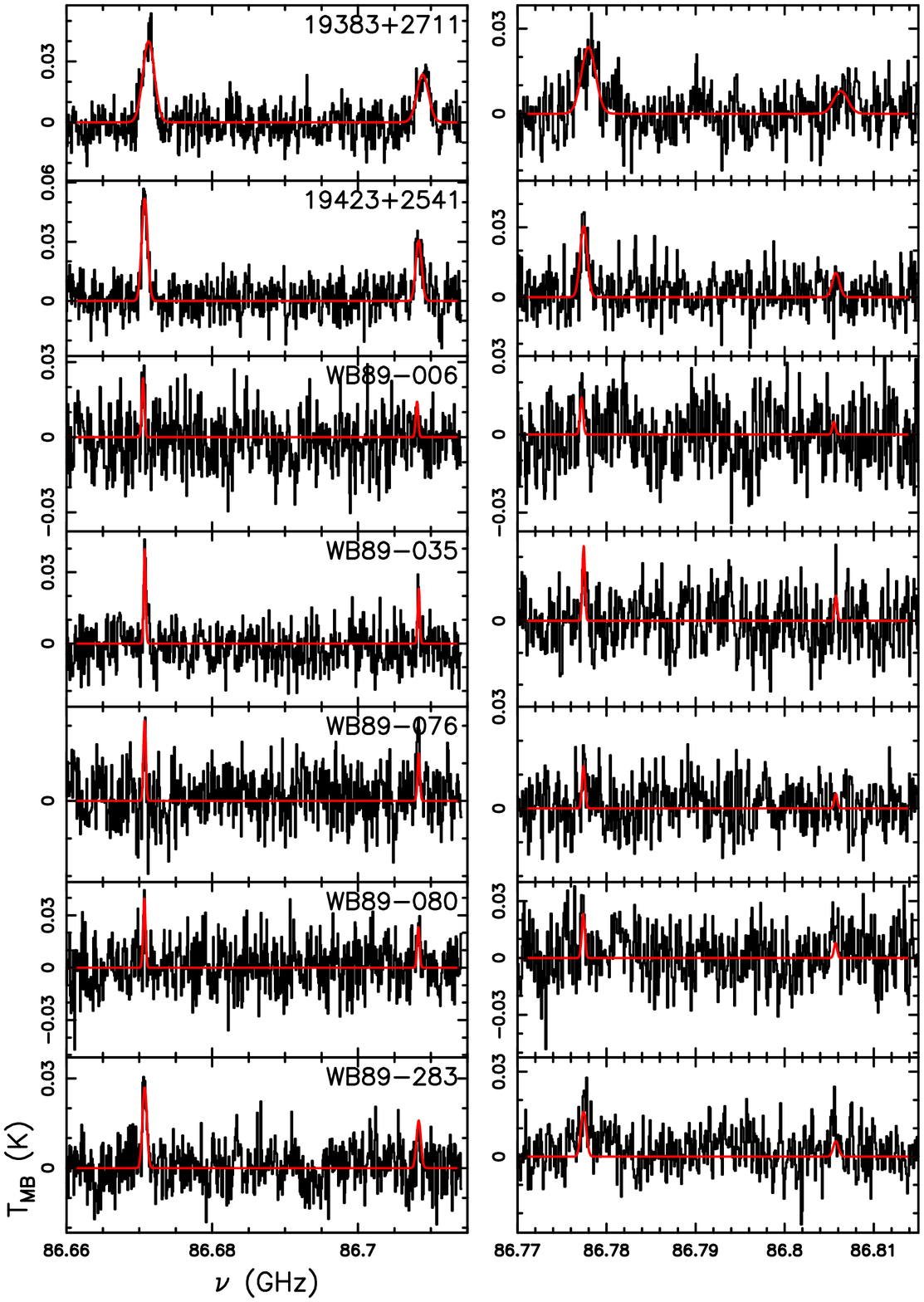}}
      \caption{Same as Fig~\ref{fig:spectra-HCO-Fig1} for the remaining seven sources.}
\label{fig:spectra-HCO-Fig2}
\end{figure*}

\FloatBarrier
\begin{figure*}
{\includegraphics[width=15cm]{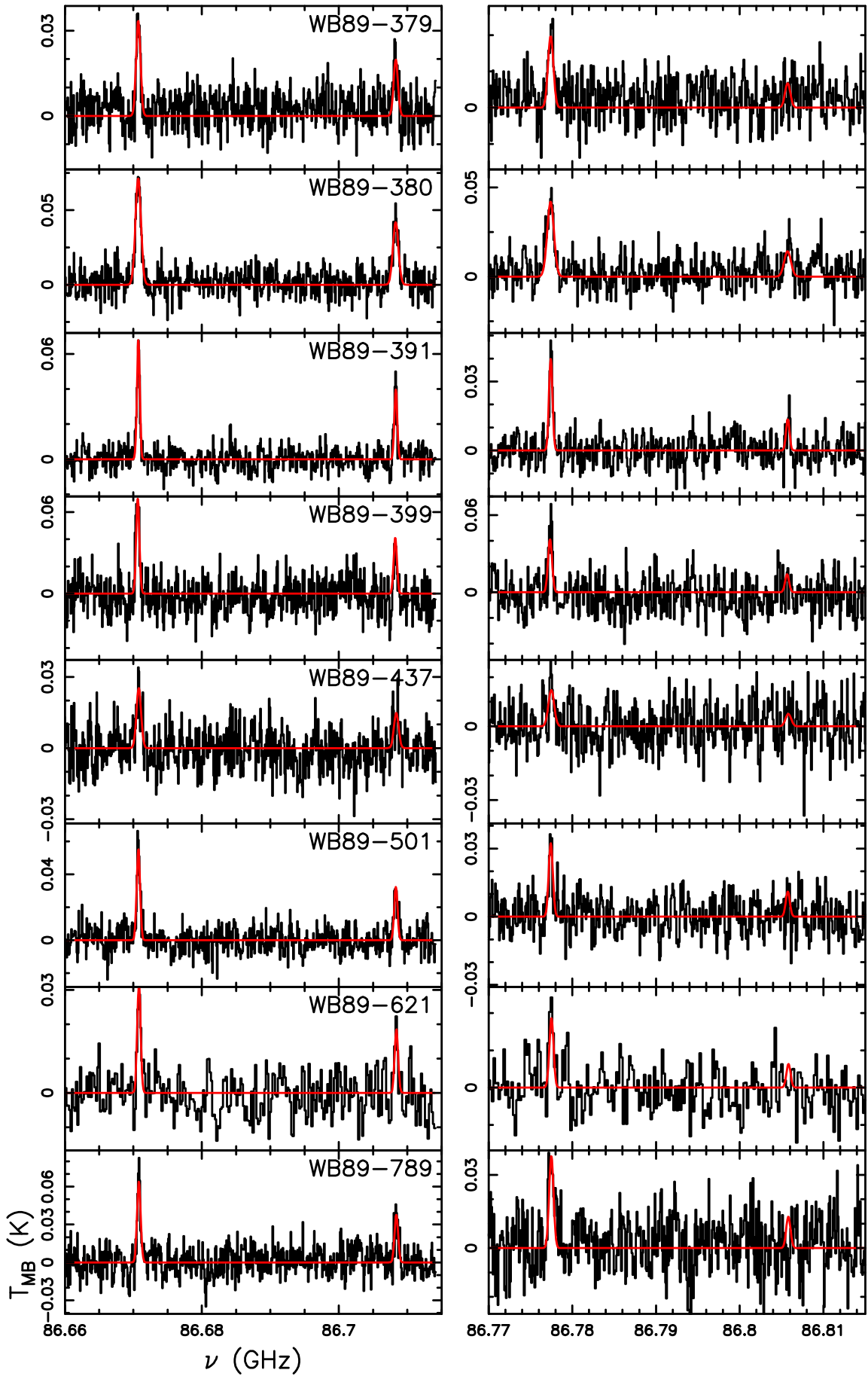}}
      \caption{Same as Fig.~\ref{fig:spectra-HCO-Fig1}, with the best fit obtained with {\sc madcuba} 
      fixing \Tex\ at the excitation temperature of \FORM\ (see Sects.~\ref{analysis} and \ref{hco}).}
\label{fig:spectra-HCO-Fig1-h2co}
\end{figure*}

\FloatBarrier
\begin{figure*}
{\includegraphics[width=15cm]{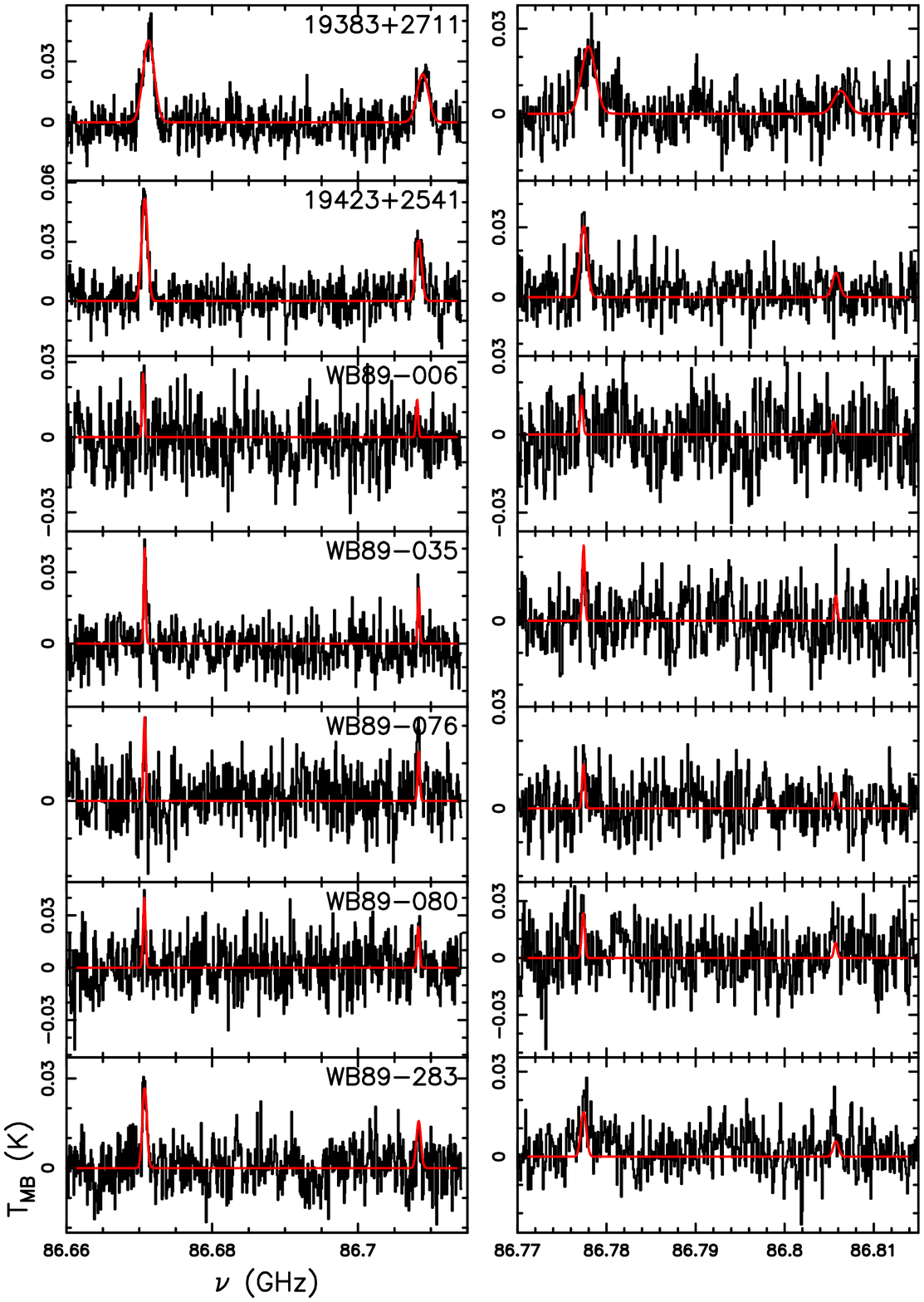}}
      \caption{Same as Fig~\ref{fig:spectra-HCO-Fig1-h2co} for the remaining seven sources.}
\label{fig:spectra-HCO-Fig2-h2co}
\end{figure*}

\FloatBarrier
\section{Results of HCO line fitting assuming \Tex\ from \FORM.}
%\section{Tables}
\label{appb}

Fit results to the HCO lines obtained with {\sc madcuba} fixing \Tex\ at the excitation temperature of 
\FORM\ (see Sects.~\ref{analysis} and \ref{hco}).

\FloatBarrier
\begin{table*}[h]
\label{tab:bestfit-hco-h2co}
\begin{center}
\caption[]{HCO line parameters. }
\begin{tabular}{ccccccc}
\hline \hline
     source        & $V$$^{(1)}$   & FWHM$^{(1)}$      & $N_{\rm tot}$$^{(1)}$       & $T_{\rm ex}$$^{(2)}$ & $X_{\rm CO}$[HCO]$^{(3)}$        & $X_{\rm Her}$[HCO]$^{(4)}$  \\
              & \kms\ & \kms      & $\times 10^{12}$ \cmq\  & K                    & $\times 10^{-10}$ & $\times 10^{-10}$ \\
\hline
    WB89-379 & --89.17(0.09) & 2.6(0.2)  &   7.1(0.5)  & 31 & 11(2)  & 11(2) \\           
    WB89-380 & --86.46(0.06) & 3.2(0.1) &   17.0(0.6)  & 28   & 15(2) & --       \\
    WB89-391 & --85.94(0.04) & 1.7(0.1)  &   7.1(0.4)  & 25  & 14(2) & --       \\
    WB89-399 & --81.79(0.09) & 2.0(0.2)  &  9.0(0.8)  & 26  & 14(2)  & 3.6(0.6) \\
    WB89-437 & --71.8(0.2)    & 2.8(0.4)  &   6.2(0.8)  & 33 & 4.5(1.0) & --       \\
    WB89-501 & --58.33(0.07) & 2.0(0.2)  &   9.8(0.7)  & 33  & 9(2) & --       \\
    WB89-621 & --25.5(0.1)   & 2.0(0.2)  &   4.2(0.5)  & 26  & 3.2(0.7) & 4.0(0.9) \\
    WB89-789 & 34.21(0.08)  & 2.1(0.2)  &   18(1.4)  & 45    & 28(5) & 16(3) \\
  19383+2711 & --68.6(0.2)   & 6.5(0.3)  &   25(1.2)  & 36  & --       & --       \\
  19423+2541 & --72.54(0.09) & 3.3(0.2)  &  19(1)  & 40   & --       & 27(4) \\
    WB89-006 & --90.5(0.1)    & 1.4(0.3)  &   2.0(0.4)  & 26  & 3.1(0.8) & --       \\
    WB89-035 & --77.60(0.07) & 1.15(0.16) &  3.0(0.4)  & 32 & 6(1) & 6(1) \\
    WB89-076 & --97.1(0.1)  & 1.4(0.2)  &   2.0(0.3)  & 28  & 4.3(0.9)   & 7(2) \\
    WB89-080 & --74.0(0.1)  & 1.4(0.3)  &   4.3(0.7)  & 30  & 5(1)   & 9(2) \\
    WB89-283 & --94.4(0.14) & 2.5(0.3)  &  6.5(0.7)  & 35  & 10(2)   & 42(10) \\         
\hline
\end{tabular}
\end{center}
$^{(1)}$ Best fit parameters obtained with {\sc madcuba}. We assumed that the emission fills the telescope beam; \\
$^{(2)}$ fixed to the value obtained from \FORM\ (Table~\ref{tab:bestfit-h2co}); \\
$^{(3)}$ fractional abundance w.r.t. H$_2$ from $N_{\rm CO}$(H$_2$), given in Table~\ref{tab:sources}; \\
$^{(4)}$ fractional abundance w.r.t. H$_2$ from $N_{\rm Her}$(H$_2$), given in Table~\ref{tab:sources}. \\
\end{table*}
\FloatBarrier

\end{appendix}

\twocolumn

\end{document}